\documentclass[11pt, a4paper]{article}
\pdfoutput=1
\usepackage{hyperref, url}
\usepackage{amssymb,amsmath,amsfonts, mathtools, mathrsfs, doi, enumerate, cite}
\usepackage[utf8]{inputenc}
\usepackage[dvipsnames]{xcolor}
\usepackage{graphicx}
\usepackage{caption}
\usepackage{tensor}
\usepackage{subcaption}
\usepackage{enumerate}
\usepackage{dsfont}
\usepackage{comment}

\usepackage{verbatim}
\usepackage{amsfonts,euscript,amssymb,braket}
\usepackage{tikz}
\usetikzlibrary{arrows,decorations.markings,patterns}
\usepackage{slashed}

\def\clock{{\count0=\time
		\divide\count0 60
		\ifnum\count0<10 0\fi\the\count0
		\multiply\count0 -60 \advance\count0 \time
		:\ifnum\count0<10 0\fi \the\count0
}}
\newcommand{\timestamp}{{\small\vbox{\hbox{\tt\jobname.tex}
			\hbox{\the\day/\the\month/\the\year, \clock}}}}

\makeatletter
\let\old@startsection=\@startsection
\let\oldl@section=\l@section
\renewcommand{\@startsection}[6]{\old@startsection{#1}{#2}{#3}{#4}{#5}{#6\mathversion{bold}}}
\renewcommand{\l@section}[2]{\oldl@section{\mathversion{bold}#1}{#2}}
\makeatother

\numberwithin{equation}{section}
\begin{document}
	\renewcommand{\thefootnote}{\arabic{footnote}}

	\overfullrule=0pt
	\parskip=2pt
	\parindent=12pt
	\headheight=0in \headsep=0in \topmargin=0in \oddsidemargin=0in

	\vspace{ -3cm} \thispagestyle{empty} \vspace{-1cm}
	\begin{flushright} 
		\footnotesize
		\textcolor{red}{\phantom{print-report}}
	\end{flushright}

\begin{center}
	\vspace{1.2cm}
	{\Large\bf \mathversion{bold} More on entanglement properties of $Lif_4^{(2)}\times {S}^1\times S^5$ spacetime with string excitations}	
	
	\vspace{0.3cm} {
		Sabyasachi Maulik\footnote[1]{sabyasachi.maulik@saha.ac.in}
	}
	\vskip  0.3cm
	
	\small
	{\em
		Theory Division, Saha Institute of Nuclear Physics, A CI of Homi Bhabha National Institute,\\ 1/AF, Bidhannagar 700064, West Bengal, India.
	}
	\normalsize	
\end{center}

\begin{abstract} 
	The $Lif_{4}^{(2)} \times S^1 \times S^5$ spacetime is an exact solution of $F1-D2-D8$ configuration in type IIA supergravity and can accommodate charged excitations of the fundamental string. By gauge/gravity duality, it is related to an excited state of a non-relativistic QFT with anisotropic Lifshitz scaling symmetry. We study mutual and tripartite information and entanglement wedge cross-section in bulk gravity for boundary subsystems that are disjoint strips of very narrow width. Our work helps understand the nature of entanglement in the QFT excited state, which is in general a mixed one.
\end{abstract}

%\tableofcontents
%\vspace{0.15 cm}
%%%%%%%%%%%%%%%%%%%%%%%%%%%%%%%%%%%%%%%%%%%%%
%%%%%%%%%%%%%%%%%%%%%%%%%%%%%%%%%%%%%%%%%%%%%
\section{Introduction}

The AdS/CFT correspondence \cite{Maldacena:1997re, Witten:1998qj} has opened a new line of investigation into large $N$ strongly coupled gauge theories. Following the pioneering works by Hubeny, Rangamani, Ryu, and Takayanagi \cite{Ryu:2006bv, Ryu:2006ef, Hubeny:2007xt}, a plethora of efforts have tried to implement ideas that naturally occur in quantum information theory to further our understanding of quantum gravity, not without exciting results. In fact, it is now widely believed that the entanglement structure among the degrees of freedom of the boundary quantum field theory plays a pivotal role in the emergence of gravitational physics in one higher dimension \cite{VanRaamsdonk:2010pw, Faulkner:2013ica}.

Non-relativistic extensions of gauge/gravity duality are very much sought after as they might be fruitful for the understanding of many realistic systems accessible to present-day experiments. In this regard, well-known attempts have been made to unearth a dual gravity description of field theories with \emph{Schrödinger} \cite{Son:2008ye, Balasubramanian:2008dm, Balasubramanian:2010uk} and \emph{Lifshitz} \cite{Kachru:2008yh} symmetries. Such field theories govern many condensed matter systems, e.g. fermions at unitarity, and multicritical points in certain magnetic materials. It is a relevant exercise to borrow tools from holography and study information theoretic aspects of such theories.

In a previous work \cite{Maulik:2019qup}, we studied holographic entanglement entropy for strip-like and spherical entangling subregions in a deformed Lifshitz spacetime. The latter was proposed as the dual geometry to some zero-temperature excited state of a $(2+1)$-dimensional Lifshitz quantum field theory in \cite{Singh:2017wei, Singh:2018ibp}. Due to its short-distance divergence structure, entanglement entropy is required to be defined with an ultraviolet cutoff and happens to be a regularization scheme dependent quantity. In this work we would like to extend our previous analysis to other well-known measures of quantum entanglement. To be more specific, we study three quantities: (i) mutual information (MI), (ii) tripartite information, and (iii) entanglement wedge cross-section (EWCS) in the asymptotically Lifshitz spacetime using holography. The first two of these are defined by making appropriate linear combinations of entanglement entropies and do not suffer from any ultraviolet divergence. The third one, besides being UV finite has recently received some limelight due to it being considered a good measure of entanglement for mixed states \cite{Takayanagi:2017knl, Nguyen:2017yqw, Umemoto:2019jlz}. 

Entanglement in Lifshitz like quantum field theories has been studied extensively from many points of view, see \cite{Chakraborty:2014lfa, MohammadiMozaffar:2017nri, Parker:2017lnh, He:2017wla, Mishra:2018tzj, MohammadiMozaffar:2018vmk, Angel-Ramelli:2019nji, Angel-Ramelli:2020xvd, Mozaffar:2021nex} for some examples. It should be mentioned though that top-down constructions of holographic dual space-time with Lifshitz scaling symmetry are not very abundant. Our setup is special in this sense because it is an exact solution of type-IIA supergravity which is at zero temperature but exhibits a non-trivial space-time geometry. Thus, it presents an opportunity to investigate features of zero-temperature excited states of a Lifshitz QFT where the excitation involves change in a $U(1)$ charge using holography. Our study is a small step in this direction.

The rest of this article is organized as follows: after a brief review of the geometry we study entanglement entropy of multiple strips in section \ref{sec2}; in the same section we calculate mutual and tripartite information for two and three strips and study how they depend on the excitation of the system. We study the entanglement wedge cross section in section \ref{sec3}. In section \ref{sec_comparison} we compare our results with those in a holographic CFT. The conclusions are given in \ref{sec4}.

\section{Multiple strips on the boundary}\label{sec2}
The $a=2$ Lifshitz vacua with IR excitations is an exact solution of massive type IIA supergravity \cite{Romans:1985tz} constituted by `massive' strings, $D2$, and $D8$ branes \cite{Singh:2017wei, Singh:2018ibp}. The spacetime metric, alongwith other non-trivial field configurations are given by
\begin{equation} \label{Lif_soln}
	\begin{split}
	&ds^2 = L^2\left(-\frac{dt^2}{z^4\, h(z)} + \frac{dx_1^2+dx_2^2}{z^2} + \frac{dz^2}{z^2} + \frac{dy^2}{q^2\, h(z)} + d\Omega_5^2 \right),\\
	&e^\phi = g_0\, h^{-1/2},
	~~~~~ C_{(3)} = - \frac{1}{g_0} \frac{L^3}{z^4}dt\wedge dx_1\wedge dx_2,\\
	&B_{(2)}=  \frac{L^2}{q\, z^2}\, h^{-1}(z)\, dt\wedge dy,
	\end{split}
\end{equation}
where $h(z) = 1 + \frac{z^2}{z_I^2}$ is a harmonic function, $q$ is a free length parameter, and the parameter $z_I$ is related to the charge of the NS-NS strings \footnote{The change in the charge of the strings winding around the compact $y$-direction is proportional to $z_I^{-2}$ \cite{Singh:2018ibp, Maulik:2019qup}.}. The $h(z) = 1$ solution is an undeformed Lifshitz space-time which, by virtue of holography is dual to the ground state of a strongly coupled QFT with anisotropic Lifshitz scaling symmetry \cite{Kachru:2008yh} \[t \to \lambda^4 t, ~~~ z \to \lambda z, ~~~ x_i \to \lambda x_i, ~~~ y \to y. \]  Accordingly, we may think of the solution \eqref{Lif_soln} as a putative dual space-time to an excited state of the same QFT. The excitations in the bulk involve $g_{tt}$ and $g_{yy}$ components of the metric but leave the $\left(x_1, x_2\right)$ plane unaffected. They also induce running of the dilaton. The string B-field couples to the excitations as well. In the deep IR $\left(z \gg z_I\right)$ the solution flows to a conformally Lifshitz solution with $a=3$ \cite{Singh:2017wei, Singh:2018ibp}.

We assume that the $y$-direction is compactified over a circle of radius $r_y$. Hence, in practice we only bother about the four dimensional theory obtained after compactification.

Entanglement entropy in the above space-time geometry was studied in \cite{Maulik:2019qup}. Holographically, the entanglement entropy between a subsystem $A$ on the asymptotic boundary and its complement is given by the Ryu-Takayanagi (RT) proposal \cite{Ryu:2006bv, Ryu:2006ef, Hubeny:2007xt}
\begin{equation}
	S_E = \frac{\text{Area}\left(\gamma_A\right)}{4 G_N^{(d+1)}}, ~~~~\partial\gamma_A = \partial A,
\end{equation}
where $\gamma_A$ is a minimal area bulk co-dimension 2 hypersurface homologous to $A$, and $G_N^{(d+1)}$ is the $(d+1)$-dimensional Newton's constant.

In this work, we concentrate only on subsystems that have the geometry of a strip. A single strip of width $\ell$ may be described as \[x_1 \in \left[-\frac{\ell}{2}, \frac{\ell}{2}\right], ~~~~x_2 \in \left[0, \ell_2\right], ~~\ell_2 \gg \ell, \] the spacetime is static and we expect that the minimal area RT hypersurface obeys the translation symmetry in the $x_2$ direction. It is then the easiest to parameterize the surface by a single function $x_1 = x_1\left(z \right)$, and its area is given by the integral
\begin{equation} \label{HEE_intgrl}
	\mathcal{A_{\gamma}} = 2L^2\ell_2\int_{\epsilon}^{z_{\ast}} \frac{dz}{z^2}\, h(z)^{\frac{1}{2}} \sqrt{1 + x_1'(z)^2},
\end{equation}
where an ultraviolet cutoff $\epsilon$ is put to regulate short-distance divergence in the area, and $z_{\ast}$ is the turning point: it's the deepest point in the bulk spacetime that the Ryu-Takayanagi hypersurface can reach.
\begin{figure}[t]
	\centering
	\includegraphics[scale = 0.2]{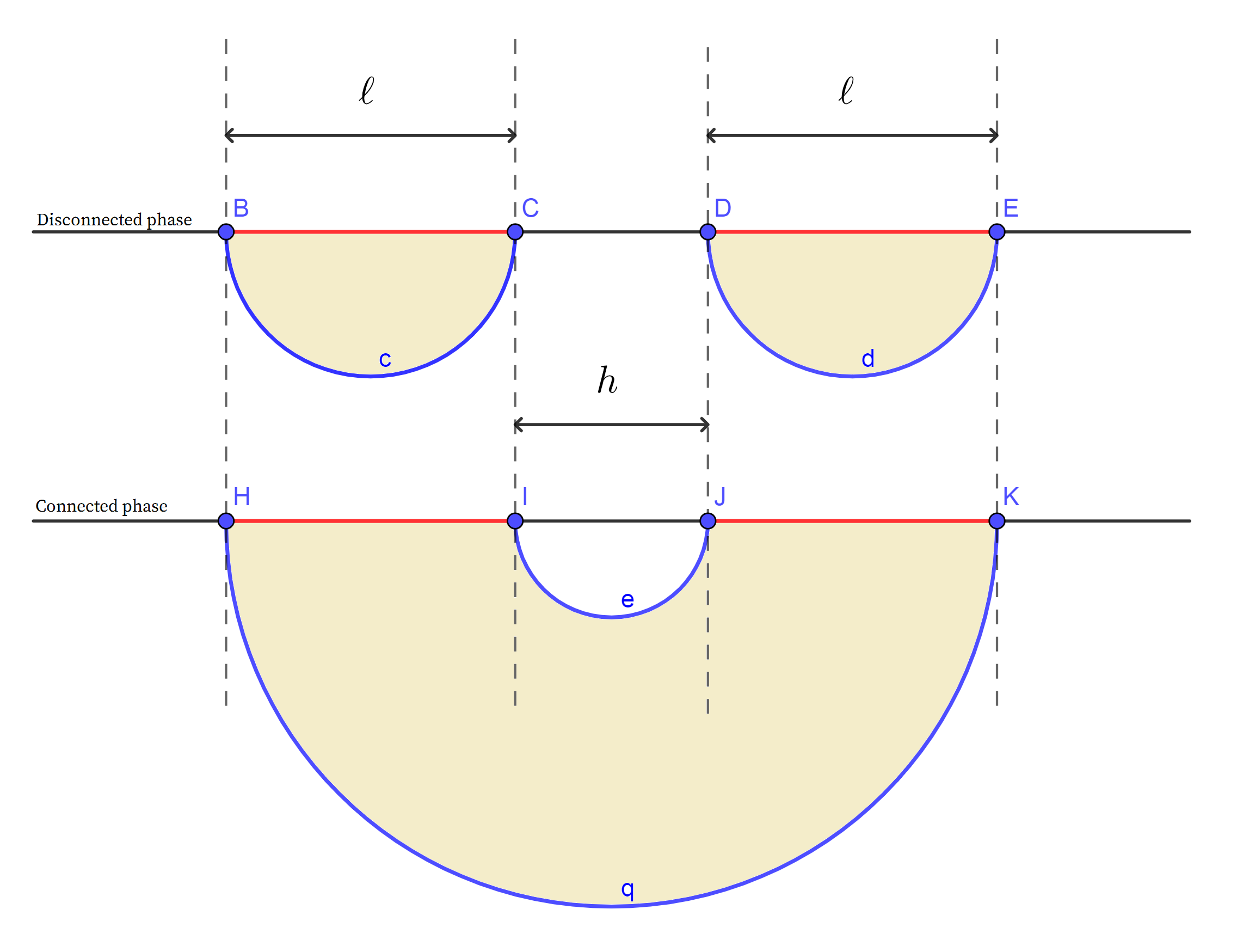}
	\caption{Two possible minimal surface configurations when the subsystem is made of two strips of equal width $\ell$ and separated by a distance $h$. The shaded area represents the entanglement wedge in the two cases. The red intervals represent the strips on the asymptotic noundary of AdS. This illustration only shows the view on the $x_1 - z$ plane.}
	\label{fig_disentangling}
\end{figure}

It is in general not possible to solve the integral analytically. Anyway, the entanglement entropy may be obtained perturbatively when $\ell \ll z_I$. Up to second order in the dimensionless perturbation parameter $\frac{\ell}{z_I}$, the turning point and the strip-width are related by
\begin{equation}\label{turnpt}
		z_* = 
		\frac{z_*^{(0)}}{1 + \frac{z_*^{(0) 2}}{2z_I^2} \frac{I_1}{b_0} - 
			\frac{z_*^{(0) 4}}{8z_I^4} \left(\frac{I_2}{b_0} + \frac{4I_1^2}{b_0^2}\right)}\,,
\end{equation}
where $b_0, I_1$ and $I_2$ are results of some integrations written in terms of Beta functions, they are
\begin{equation} \label{beta_fn_parameters}
	\begin{split}
		b_0 &= \frac{1}{4} B\left(\frac{3}{4}, \frac{1}{2}\right), \\
		I_1 &= \frac{1}{4}\left(B\left(\frac{3}{4},-\frac{1}{2}\right) - B\left(\frac{5}{4},-\frac{1}{2}\right)\right), \\
		I_2 &= \frac{1}{8}\left(2B(\frac{3}{4}, -\frac{3}{2}) - 
		3B(\frac{5}{4}, -\frac{3}{2})\right),
	\end{split}
\end{equation}
and $z_{\ast}^{(0)} = \frac{\ell}{2 b_0}$ is the turning point in the ground state.  The holographic entanglement entropy including up to second order terms in the perturbation series is then expressed as
\begin{align} \label{HEE_result}
	S_{E} = & \frac{L^2 \ell_2}{4 G_N^{(4)}} \left(\frac{2}{\epsilon} - \frac{4 b_0^2}{\ell} + \frac{a_1}{2 b_0}\frac{\ell}{z_I^2} - \frac{1}{64}\frac{\ell^3}{z_I^4}\frac{1}{b_0^2}\left(\frac{a_1^2}{b_0^2} - 1 \right) \right), \nonumber \\
	= & \frac{L^2 \ell_2}{4 G_N^{(4)}} \left(\frac{2}{\epsilon} - \frac{4 b_0^2}{\ell} + \mathcal{S}_{E}^{(2)}\left(\ell \right) \right),
\end{align}
where the coefficient $a_1 = \frac{B\left(\frac{1}{4}, \frac{1}{2}\right)}{4}$, and $G_N^{(4)}$ is the four dimensional Newton's constant: $\frac{1}{G_N^{(4)}} = \frac{2\pi L r_y}{G_N^{(5)}}$. The first two terms in the parentheses are the EE of the ground state while the next two terms (denoted collectively by $\mathcal{S}_{E}^{(2)}\left(\ell \right)$) are corrections due to charged excitations. The leading order change to HEE is found to be negative and the same at subleading order is positive, this is because the perturbation series is convergent for $\ell < z_I$. We shall witness this time and again for all the quantities calculated later.

\begin{figure}[t]
	\centering
	\includegraphics[width=0.5\textwidth]{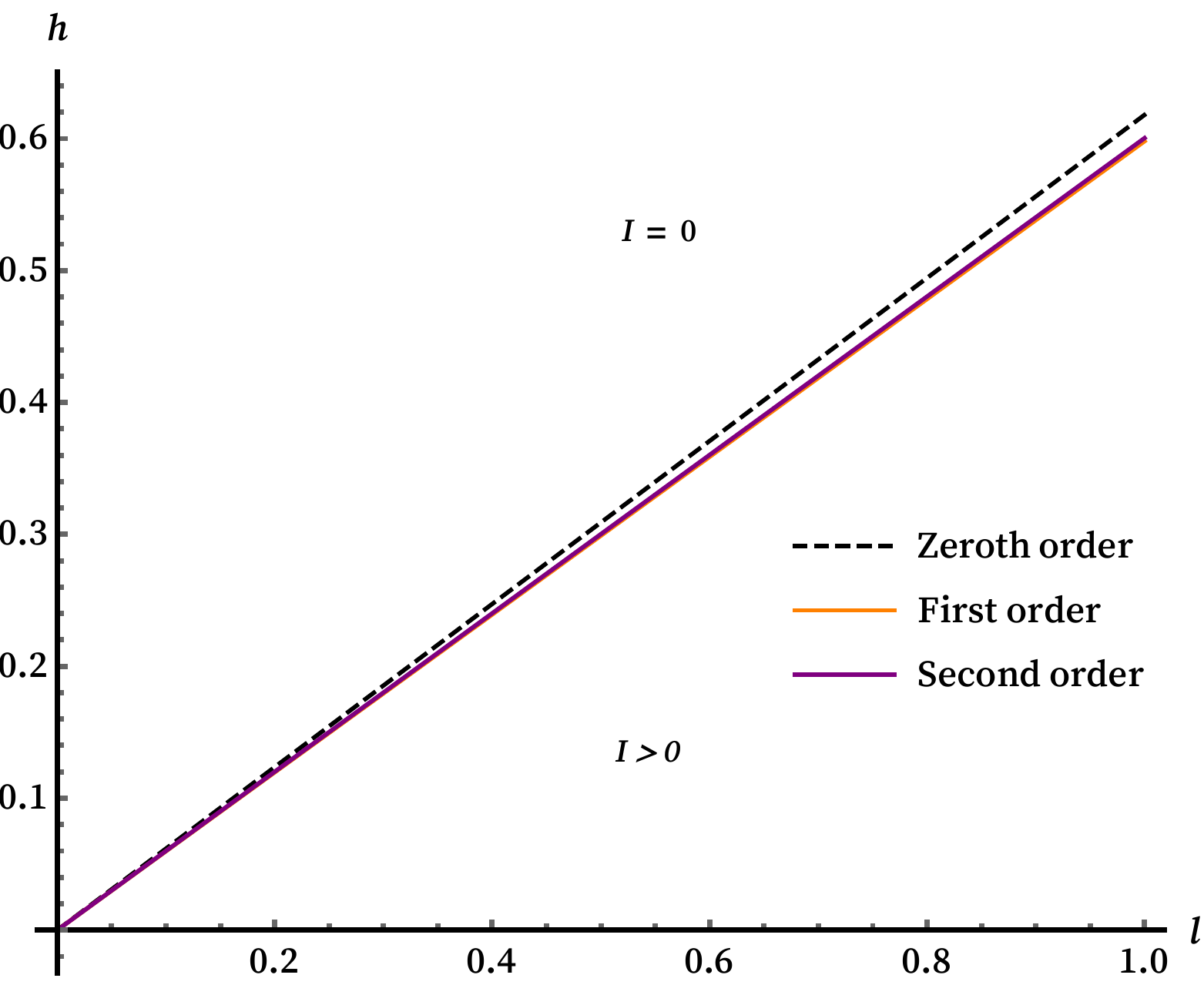}
	\caption{Lines of phase transition for $\frac{\ell}{z_I} = 0.25$}
	\label{phase_diag_eq}
\end{figure}

Now we consider two strips of equal width $\ell$ and separated by a distance $h$. For such a configuration, there are two competing Ryu-Takayanagi surfaces, schematically shown in figure \ref{fig_disentangling}. The minimum area choice between these two depends on the ratio $x = \frac{h}{\ell}$. Usually for small $x$ the `connected' hypersurface is preferred and for large $x$-values (i.e. when the two strips are sufficiently faraway from each other), the entanglement entropy is determined by the sum of two `disconnected' hypersurfaces. Thus, at a critical value $x_c$, the entanglement entropy undergoes a geometric phase transition \cite{Headrick:2010zt, Fischler:2012uv, Ben-Ami:2014gsa}. Taking this into account, we may write
\begin{align} \label{HEE_phase_transition}
	S_E\left(A \cup B\right) = \begin{dcases*}
		S_E \left(h\right) + S_E\left(h + 2\ell \right), & if $x < x_c$,\\
		2 S_E\left(\ell\right), & if $x > x_c$.
	\end{dcases*}
\end{align}
The critical point is determined by solving
\begin{equation}
	2 S_E\left(\ell\right) = S_E\left(h\right) + S_E\left(h+2\ell\right),
\end{equation}
or using eq. \eqref{HEE_result}
\begin{multline} \label{critical_length}
	\left(2 - \frac{1}{x} - \frac{1}{x+2} \right) - \frac{a_1}{8 b_0^3} \left(2 - x - \left(x+2 \right) \right) \frac{\ell^2}{z_I^2} + \frac{1}{256 b_0^4} \left( \frac{a_1^2}{b_0^2}-1 \right) \left(2 - x^3 - (x+2)^3 \right) \frac{\ell^4}{z_I^4} = 0,
\end{multline}
The solution to this equation is a straight line of phase transition in the $h-\ell$ plane \footnote{For the zeroth order or ground state, we see that the transition takes place along $\frac{h}{\ell} \approx 0.618$, this is the same for asymptotically AdS$_4$ geometry because at zeroth order (no excitation) the two extremal surface integrals are identical.}, see figure \ref{phase_diag_eq}. The true line of phase transition would lie somewhere between the black and orange curves, since our result is only perturbatively true.

\begin{figure}[t]
	\centering
	\includegraphics[width=0.5\textwidth]{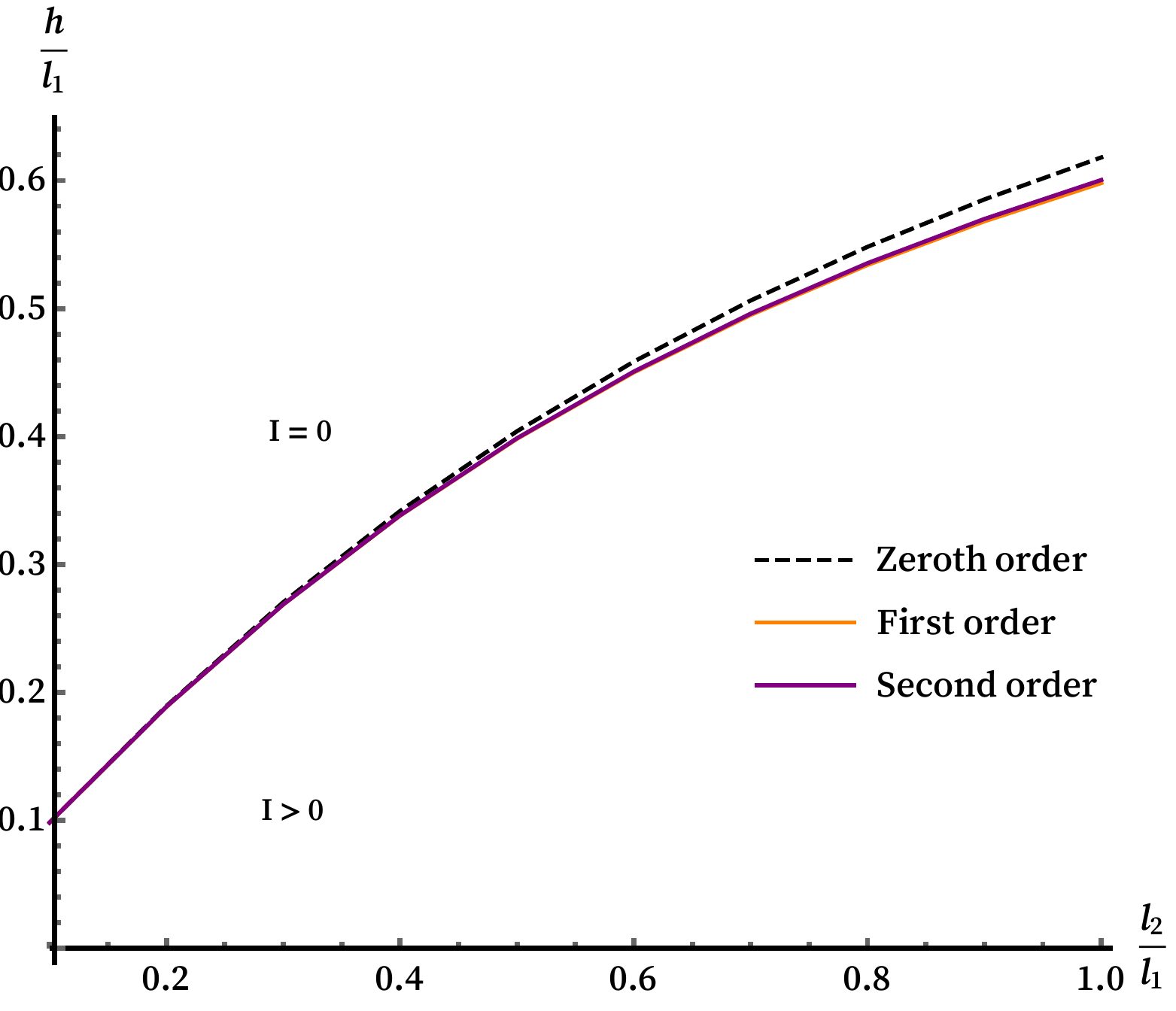}
	\caption{Lines of phase transition at different orders for two strips of widths $\ell_1$ and $\ell_2$, we choose $\frac{\ell_1}{z_I} = 0.25$.}
	\label{phase_diag_uneq}
	\hfill
\end{figure}

It is also not very hard to generalize this result for two strips of different widths $\ell_1$ and $\ell_2$. The essential features remain the same. In this case, the phase transition would take place along a curve that is the solution to
\begin{multline}
	\left(1 + \frac{1}{y} - \frac{1}{x} - \frac{1}{1 + x + y}\right) - 
	\frac{a_1}{(8 b_0^3)} \left(\frac{\ell}{z_I}\right)^2 \left(1 + y - x - \left(1 + x + y\right)\right) \\ + 
	\frac{1}{256 b_0^4}
	\left(\frac{\ell}{z_I}\right)^4 \left(\frac{a_1^2}{b0^2} - 1 \right) \left(1 + y^3 - 
	x^3 - \left(1 + x + y\right)^3 \right) = 0,
\end{multline}
where $x = \frac{h}{\ell_1}$, and $y = \frac{\ell_2}{\ell_1}$. The phase diagram for two strips of unequal widths is shown in figure \ref{phase_diag_uneq}.

\subsection{Mutual and tripartite information}
The mutual information between two disjoint regions $A$ and $B$
\begin{equation} \label{MI_def}
	I\left(A:B\right) = S_E\left(A\right) + S_E\left(B\right) - S_E\left(A \cup B\right),
\end{equation} 
measures the total amount of correlation (both classical and quantum) between $A$ and $B$. By construction, $I\left(A:B \right)$ happens to be finite and independent of any UV regulator. The mutual information is always non-negative because of the subadditivity property of entanglement entropy. \footnote{The reader is referred to \cite{Alishahiha:2014jxa, Tanhayi:2015cax, Mirabi:2016elb} for some interesting works on mutual and n-partite information.}

We have discussed how the entanglement entropy shows a crossover from a connected phase to a disconnected phase as $\frac{h}{\ell}$ is varied. From the definition of mutual information and the equations \eqref{HEE_result}, \eqref{HEE_phase_transition}  we can easily reason that
\begin{align} \label{MI_expression1}
	I\left(A:B\right) = \begin{dcases*}
		\mathcal{I}_{AB}, & if $ x < x_c $, \\
		0, & if $ x > x_c $,
	\end{dcases*}
\end{align}
therefore, mutual information undergoes a first order phase transition at $x = x_c$. Here $\mathcal{I}_{AB}$ is a finite positive quantity given by
\begin{equation} \label{MI_expression2}
	\mathcal{I}_{AB} = \frac{L^2 \ell_2}{4 G_N^{(4)}}\times \frac{4 b_0^2}{\ell} \left(\frac{\ell}{h} + \frac{1}{2 + \frac{h}{\ell}} - 2 + 2\mathcal{S}_E^{(2)}\left(\ell \right) - \mathcal{S}_E^{(2)}\left(h \right) - \mathcal{S}_E^{(2)}\left(h + 2\ell \right) \right),
\end{equation}

\begin{figure}[t]
	\begin{subfigure}{0.48\textwidth}
		\centering
		\includegraphics[width=\textwidth]{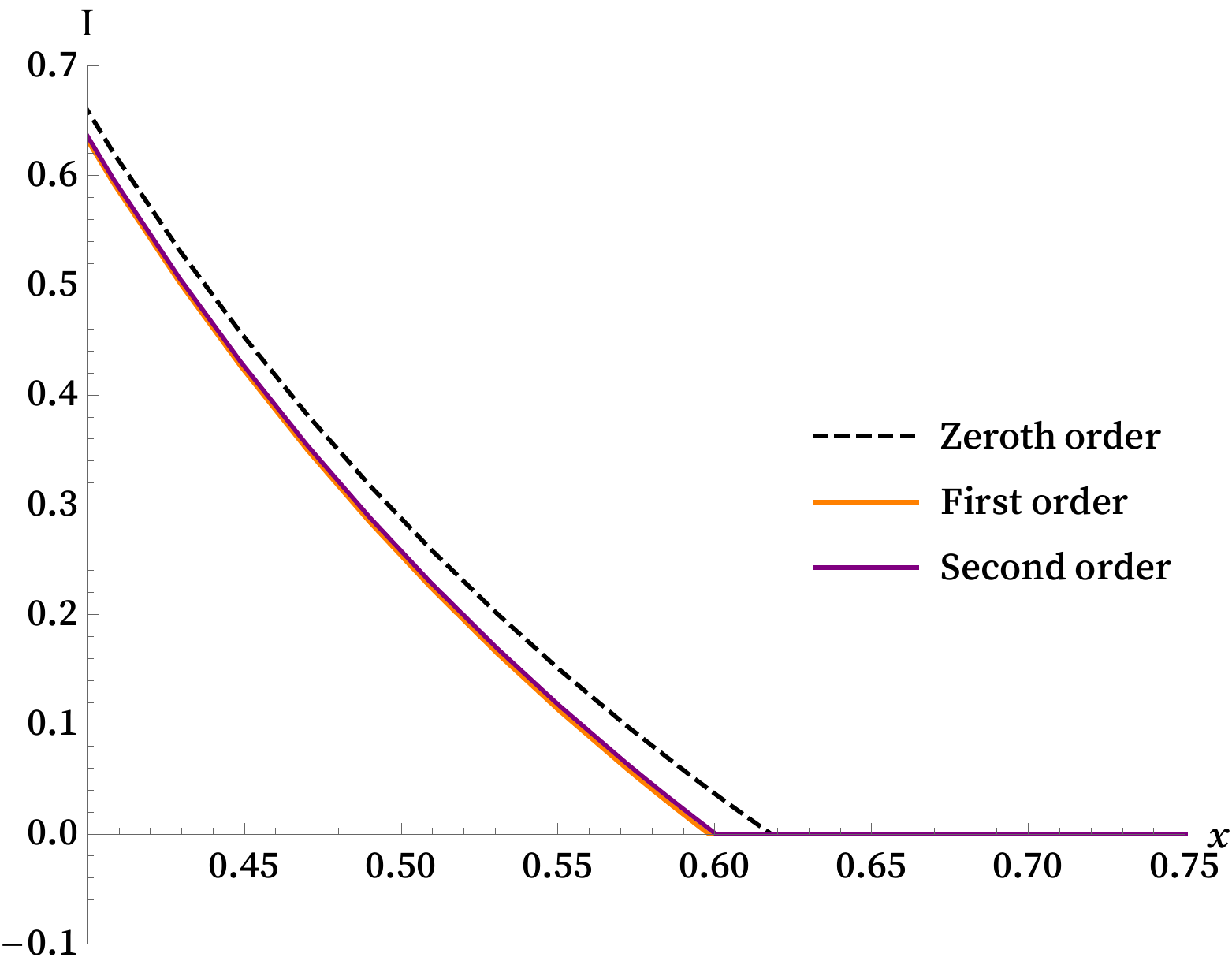}
		\caption{}
		\label{fig_MI_diff_order}
	\end{subfigure}
	\hfill
	\begin{subfigure}{0.48\textwidth}
		\centering
		\includegraphics[width=\textwidth]{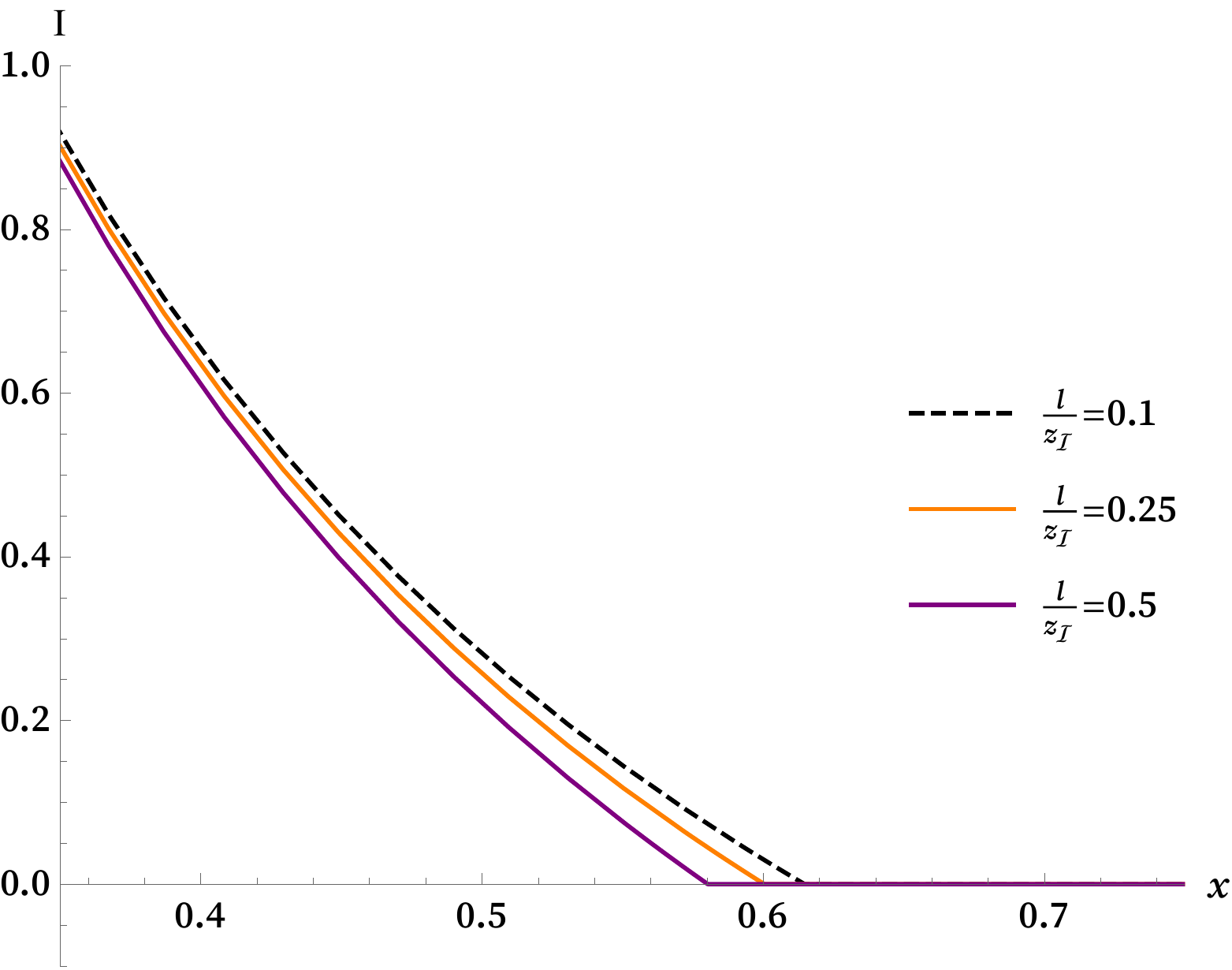}
		\caption{}
		\label{MI_diff_m}
	\end{subfigure}
	\caption{Mutual information of two strips of equal width for (a) different orders in perturbation theory with $\frac{\ell}{z_I} = 0.5$, and (b) different $\frac{\ell}{z_I}$ ratio. The $y$-axis is $\ell \times I(A: B)$ and we choose $\frac{L^2 \ell_2}{2 G_N^{(4)}} = 1$.}
	\label{MI_eq}
\end{figure}

The mutual information between two strips of equal width $\ell$ and separation $h$ in the geometry \eqref{Lif_soln} is shown in figure \ref{MI_eq}, the X-axes of which is the dimensionless ratio $x = \frac{h}{\ell}$. The plots illustrate all the expected features. We observe that the system is in its connected phase when $x$ is small and $I\left(A:B\right) > 0$. The mutual information continues to decrease monotonically as $x$ increases, and finally reaches zero when the system enters the disconnected phase, where $S\left(A \cup B\right) = S(A) + S(B)$. The figure \ref{fig_MI_diff_order} displays that the leading order change to holographic mutual information is negative and the change at subleading order is positive, this is again due to convergence of the perturbation series as we have mentioned earlier.

From figure \ref{MI_diff_m} we also observe that the mutual information between two strips decreases as the ratio $\frac{\ell}{z_I}$ is increased. Here, let us remind ourselves that the parameter $z_I$ is related to the charge of the fundamental strings that wrap around the periodic $y$-direction.  In fact, the net change in the charge density of the winding strings due to the excitations is \cite{Singh:2018ibp, Maulik:2019qup}
\begin{equation}
	\triangle \rho = \frac{2 L}{G_N^{(5)} z_I^2}.
\end{equation}
We conclude that the decrease in MI happens due to decrease in correlation between the two subregions as the degree of excitation in the system increases (assuming that their width remains unchanged). This is similar in spirit to the behaviour addressed in \cite{Fischler:2012uv}, where the MI was found to decrease due to increase in temperature of a relativistic CFT, ours is of course a zero temperature non-relativistic system.

\begin{figure}[t]
	\centering
	\begin{subfigure}{0.48\textwidth}
		\centering
		\includegraphics[width=\textwidth]{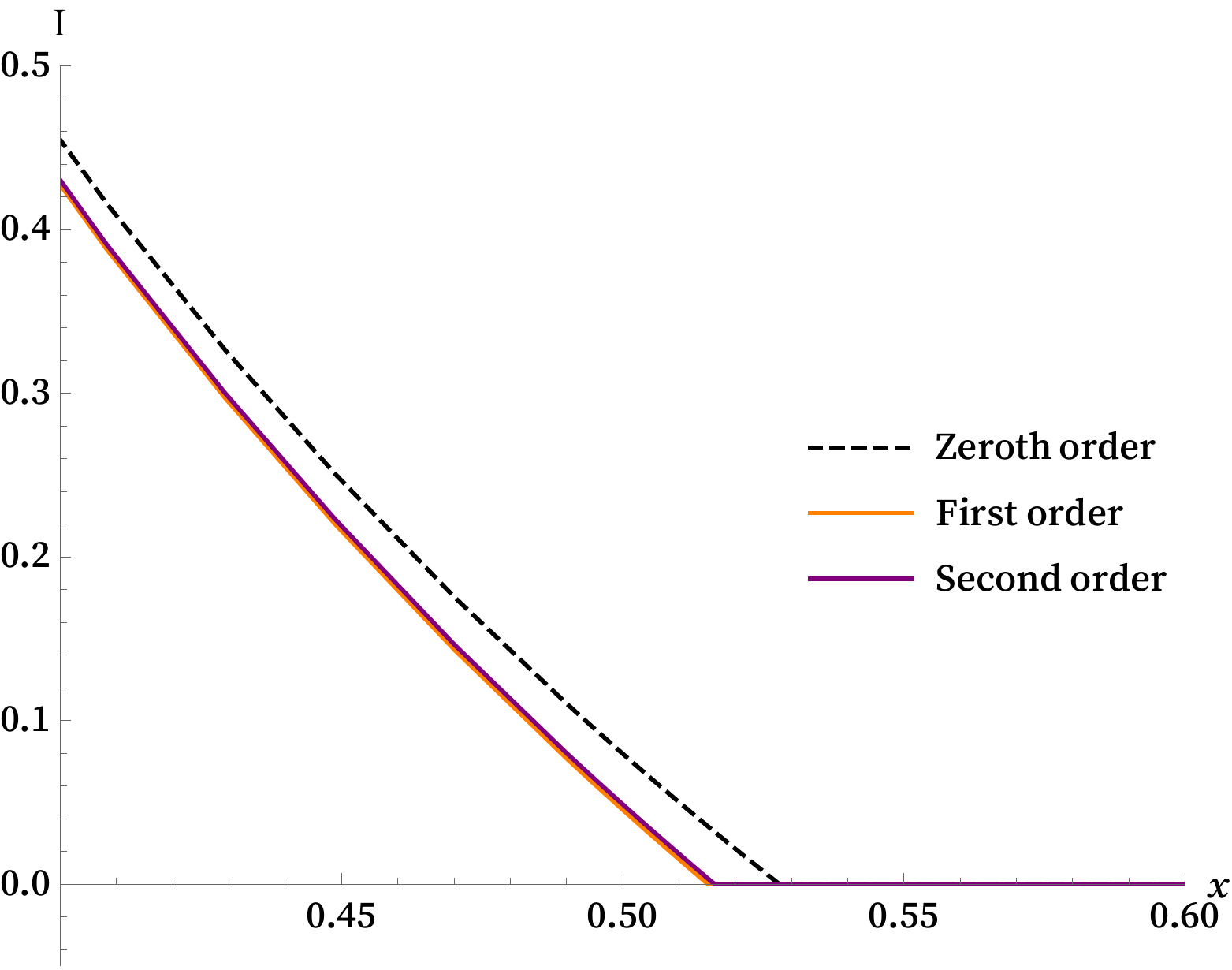}
		\caption{}
	\end{subfigure}
	\hfill
	\begin{subfigure}{0.48\textwidth}
		\centering
		\includegraphics[width=\textwidth]{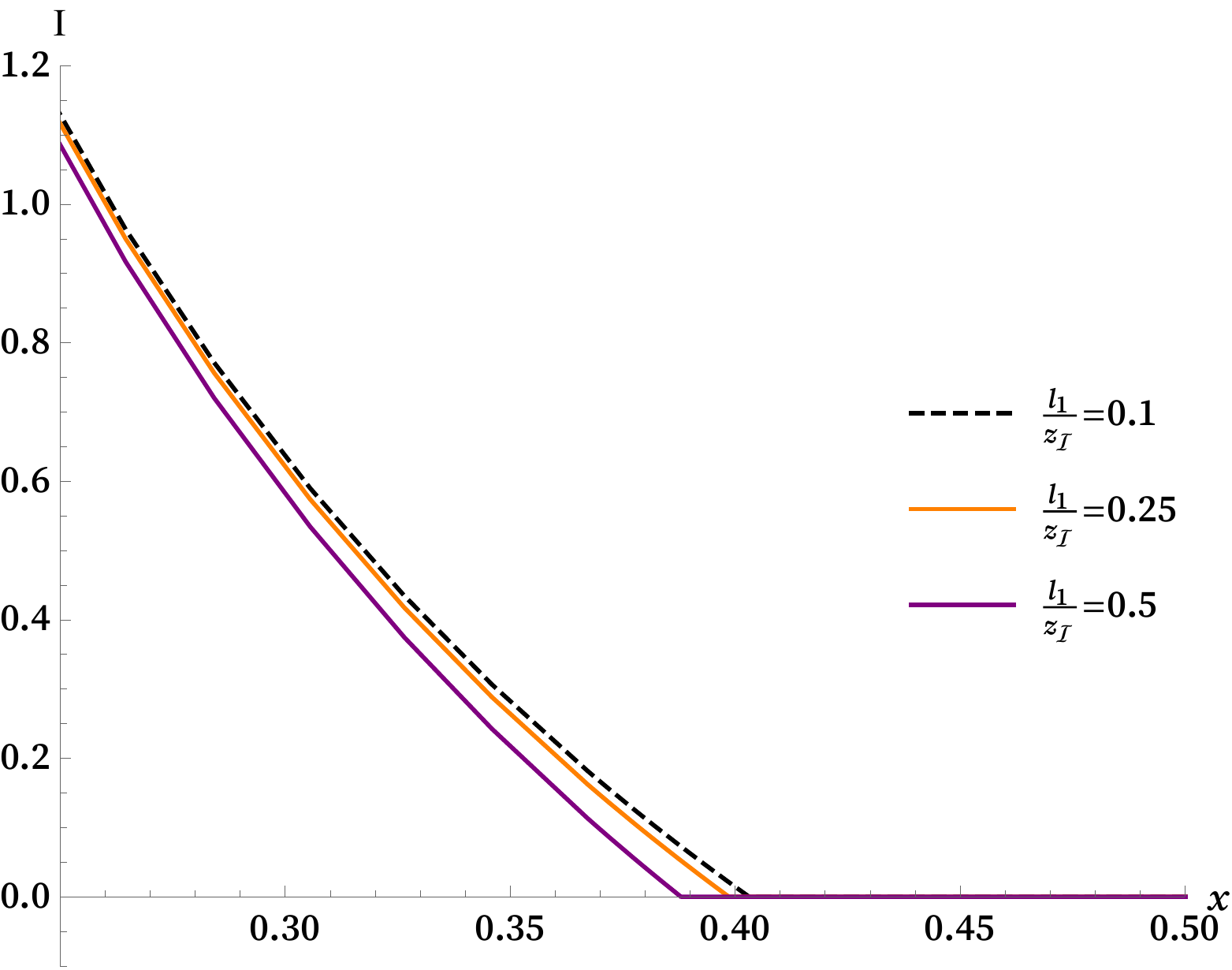}
		\caption{}
	\end{subfigure}
	\hfill
	\begin{subfigure}{0.48\textwidth}
		\centering
		\includegraphics[width=\textwidth]{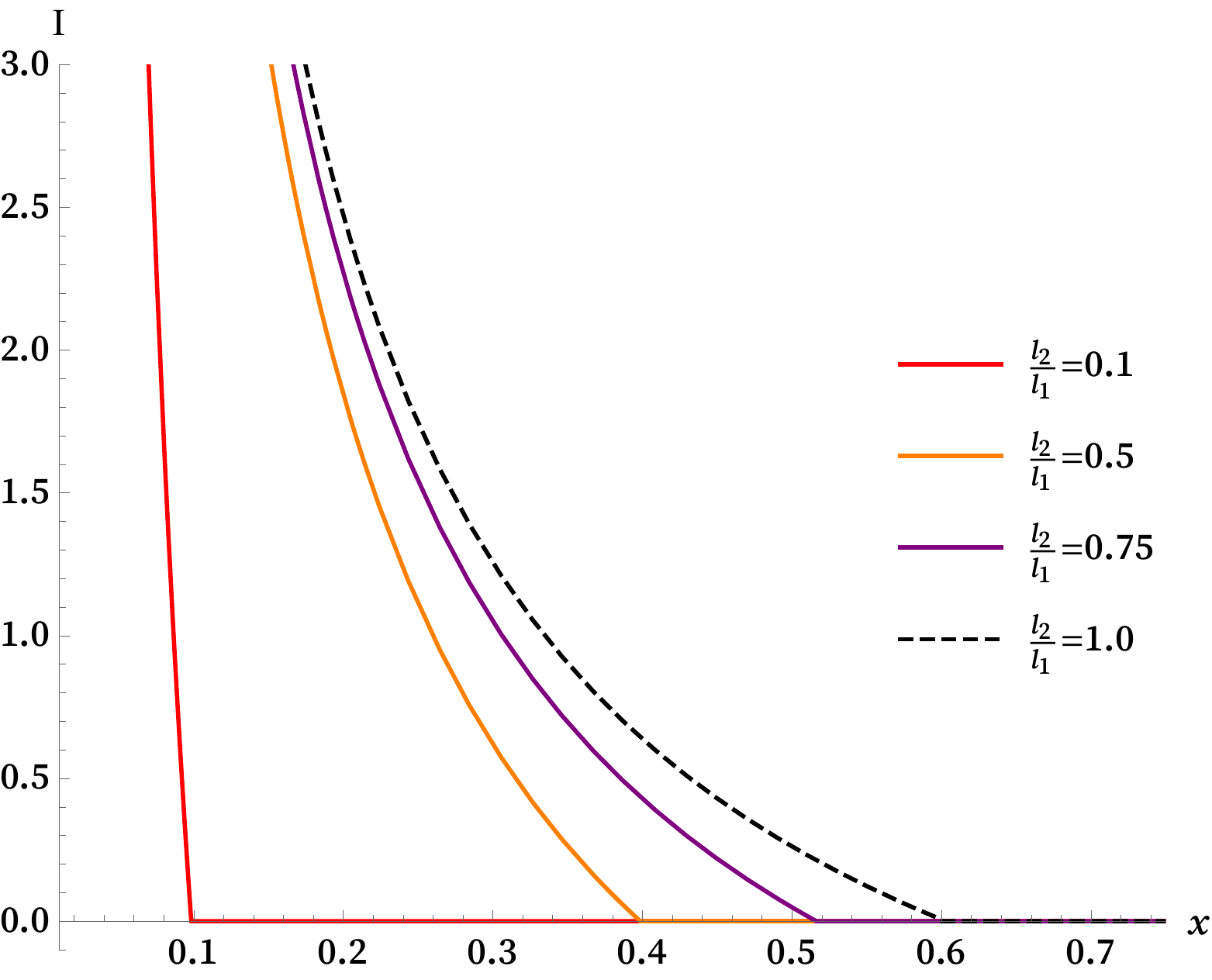}
		\caption{}
	\end{subfigure}
	\caption{Mutual information of two strips of different widths (a) for different orders in perturbation theory when $\frac{\ell_1}{z_I} = 0.25$ and $\frac{\ell_2}{\ell_1} = 0.75$, (b) for different $\frac{\ell_1}{z_I}$ ratio, where $\frac{\ell_2}{\ell_1} = 0.5$, and (c) for different $\frac{\ell_2}{\ell_1}$ ratio, where $\frac{\ell_1}{z_I} = 0.25$. The $y$-axis is $\ell \times I(A: B)$, we choose $\frac{L^2 \ell_2}{2 G_N^{(4)}} = 1$.}
	\label{MI_uneq}
\end{figure}

Mutual information between two strips of unequal widths can be similarly studied. The results are illustrated in figure \ref{MI_uneq}. The qualitative behaviour with change of separation length unsurprisingly remains the same and again MI is observed to decrease with increase of excitation. We also notice that among the class of two strips, mutual information is maximum for the one with equal width. 

\subsubsection{Tripartite information} \label{subsec_tpi}
A more complicated linear combination of entanglement entropies is the tripartite information
\begin{equation} \label{tripartite_defn}
	I_{3}\left(A: B: C\right) = I\left(A:B\right) + I\left(A:C\right) - I\left(A: B \cup C\right),
\end{equation}
where $A$, $B$, and $C$ are three disjoint regions. The tripartite information function may be thought as a measure of information shared between $B$ and $C$ with respect to $A$ \cite{Casini:2008wt}. It is also a measure of `extensivity' of mutual information. According to the Ryu-Takayanagi proposal, mutual information for any choice of regions is either extensive or super-extensive in holographic theories, which in turn implies that the tripartite information must be negative or zero \cite{Hayden:2011ag}
\begin{equation} \label{tripartite_inequality}
	I_{3}\left(A: B: C\right) \leq 0.
\end{equation}

\begin{figure}[t]
	\centering
	\includegraphics[width=\linewidth]{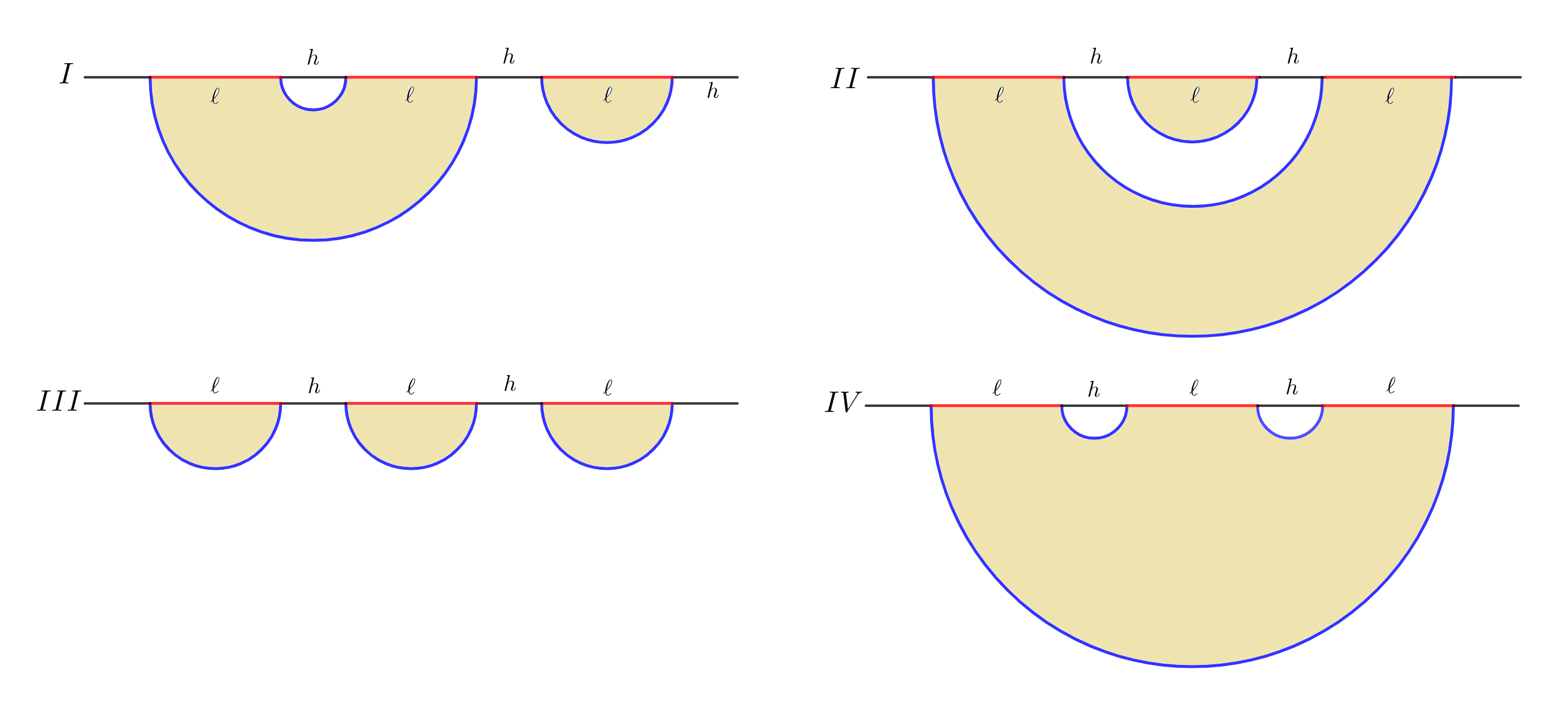}
	\caption{Candidate minimal surface configurations to determine $S\left(A: B: C\right)$ for three strips of equal width and equal separation.}
	\label{fig_tripartite_possibilities}
\end{figure}

We study holographic tripartite information in our asymptotically Lifshitz setup for three disjoint strips of equal width $\ell$ and equal separation $h$. For the case of three strips, there are multiple possible configurations for the Ryu-Takayanagi hypersurface. But we find that in the region of validity of our perturbative analysis only two of them (III and IV in figure \ref{fig_tripartite_possibilities}) would contribute. In fact, candidates I and II from the same figure are not minimal as long as the two separation distances remain equal \cite{Ben-Ami:2014gsa}.

We can use our previously obtained results \eqref{HEE_result}, \eqref{MI_expression1} and \eqref{MI_expression2} and find an analytic expression of $I_3$ valid in the perturbative regime $\ell \ll z_I$. It is easier to write the three pieces on the r.h.s. of equation \eqref{tripartite_defn} separately
\begin{align} \label{MI_expression1}
	I\left(A:B\right) & = \begin{dcases*}
		\frac{2G_N^{(4)}}{L^2\ell_2}\times\ell\times\mathcal{I}_{AB}^{(1)}, & if $ x < x_{c_1} $, \\
		0, & if $ x > x_{c_1} $.
	\end{dcases*}\\
	I\left(A:C\right) & = 0, \text{ for all } x.
\end{align}
\begin{align}
	& I\left(A: B \cup C\right) = \begin{dcases*}
		\frac{2G_N^{(4)}}{L^2\ell_2}\times\ell\times\mathcal{I}_{ABC}^{(1)}, & if $ x < x_{c_1} $, \\
		\frac{2G_N^{(4)}}{L^2\ell_2}\times\ell\times\mathcal{I}_{ABC}^{(2)}, & if $ x_{c_1} < x < x_{c_2} $,\\
		0, & if $ x > x_{c_2} $.
	\end{dcases*}
\end{align}
Where the non-zero pieces are given by
\begin{align}
	\begin{split}
		\mathcal{I}_{AB}^{(1)} = 2 \left( -\frac{1}{128\,b_0^2}\left(\frac{a_1^2}{b_0^2}-1\right)\frac{\ell^4}{z_I^4} + \frac{a_1}{4\, b_0}\frac{\ell^2}{z_I^2} - 2 b_0^2 \right) - \left(-\frac{1}{128\, b_0^2} \left(\frac{a_1^2}{b_0^2}-1 \right) \left(x^3+(x+2)^3\right)\frac{\ell^4}{z_I^4} \right. \\ \left. + \frac{a_1}{4\, b_0} \left(\left(x+2\right)+x\right)\frac{\ell^2}{z_I^2} - 2 b_0^2 \left(\frac{1}{x}+\frac{1}{x+2}\right) \right),
	\end{split} \nonumber
	\\
	\begin{split}
		\mathcal{I}_{ABC}^{(1)} = \left( -2 b_0^2 + \frac{a_1}{4\,b_0}\frac{\ell^2}{z_I^2} - \frac{1}{128\,b_0^2}\left(\frac{a_1^2}{b_0^2} - 1 \right) \frac{\ell^4}{z_I^4} \right) + \left(-2 b_0^2 \left(\frac{1}{x} + \frac{1}{x+2} \right) + \frac{a_1}{4\,b_0}\left(x + \left(x+2\right)\right)\frac{\ell^2}{z_I^2} \right. \\ \left. - \frac{1}{128\,b_0^2}\left(\frac{a_1^2}{b_0^2} - 1 \right) \left(x^3 + \left(x+2\right)^3 \right)\frac{\ell^4}{z_I^4} \right),
	\end{split} \nonumber
	\\
	\begin{split}
		\mathcal{I}_{ABC}^{(2)} = 3 \left(-2 b_0^2 + \frac{a_1}{4\,b_0}\frac{\ell^2}{z_I^2} - \frac{1}{128\,b_0^2}\left(\frac{a_1^2}{b_0^2} - 1 \right) \frac{\ell^4}{z_I^4} \right) - \left(-2 b_0^2 \left(\frac{2}{x} + \frac{1}{2x+3} \right) + \frac{a_1}{4\,b_0}\left(x + x + \left(2x+3\right)\right)\frac{\ell^2}{z_I^2} \right. \\ \left. - \frac{1}{128\,b_0^2}\left(\frac{a_1^2}{b_0^2} - 1 \right) \left(2x^3 + \left(2x+3\right)^3 \right)\frac{\ell^4}{z_I^4} \right),
	\end{split} \nonumber
 \end{align}
and the two critical points $x_{c_1}$ and $x_{c_2}$ are roots of $\mathcal{I}_{AB}^{(1)} = 0, \text{ and } \mathcal{I}_{ABC}^{(2)} = 0,$ respectively.

\begin{figure}[t]
	\begin{subfigure}{0.48\textwidth}
		\centering
		\includegraphics[width=\textwidth]{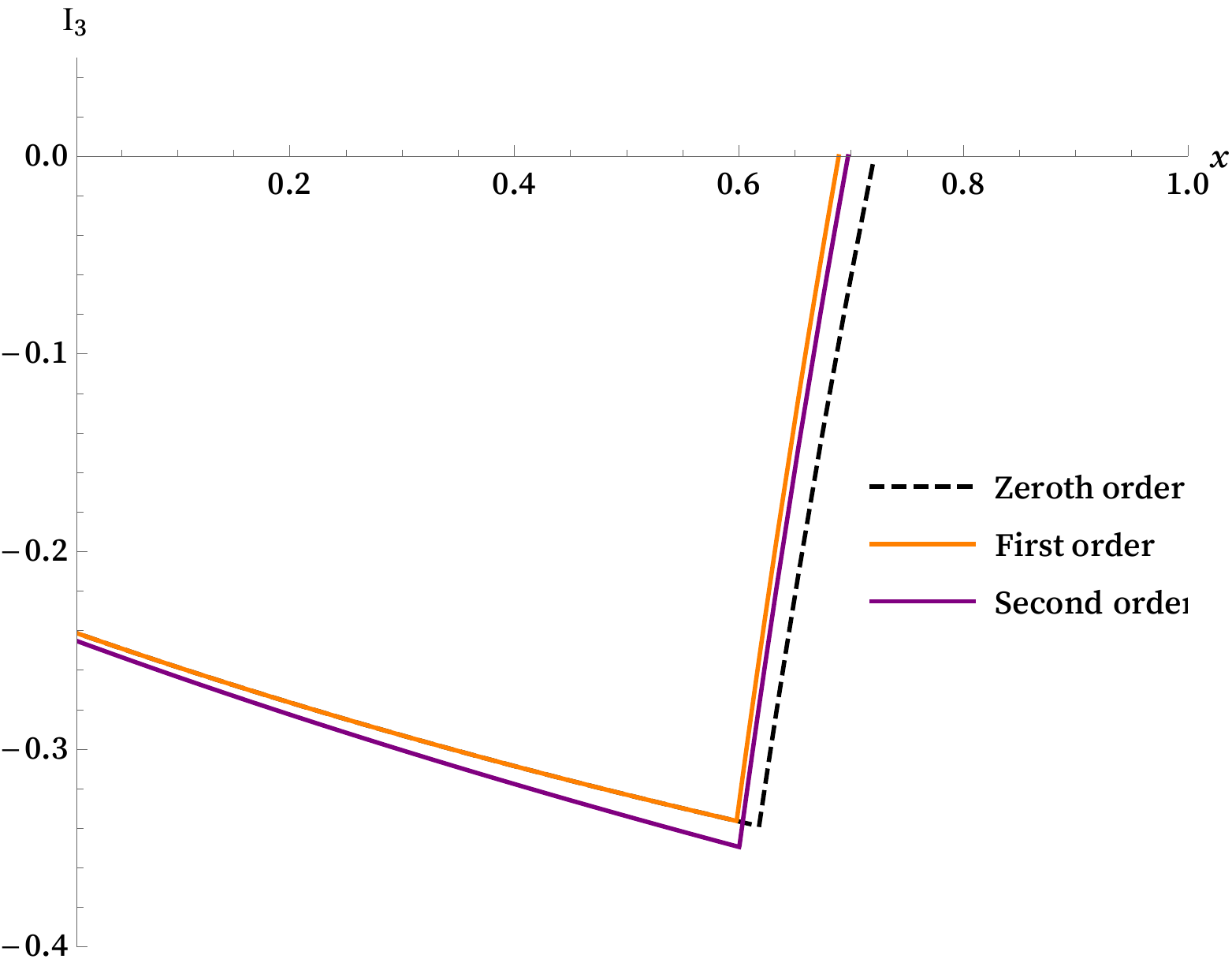}
		\caption{}
	\end{subfigure}
	\hfill
	\begin{subfigure}{0.48\textwidth}
		\centering
		\includegraphics[width=\textwidth]{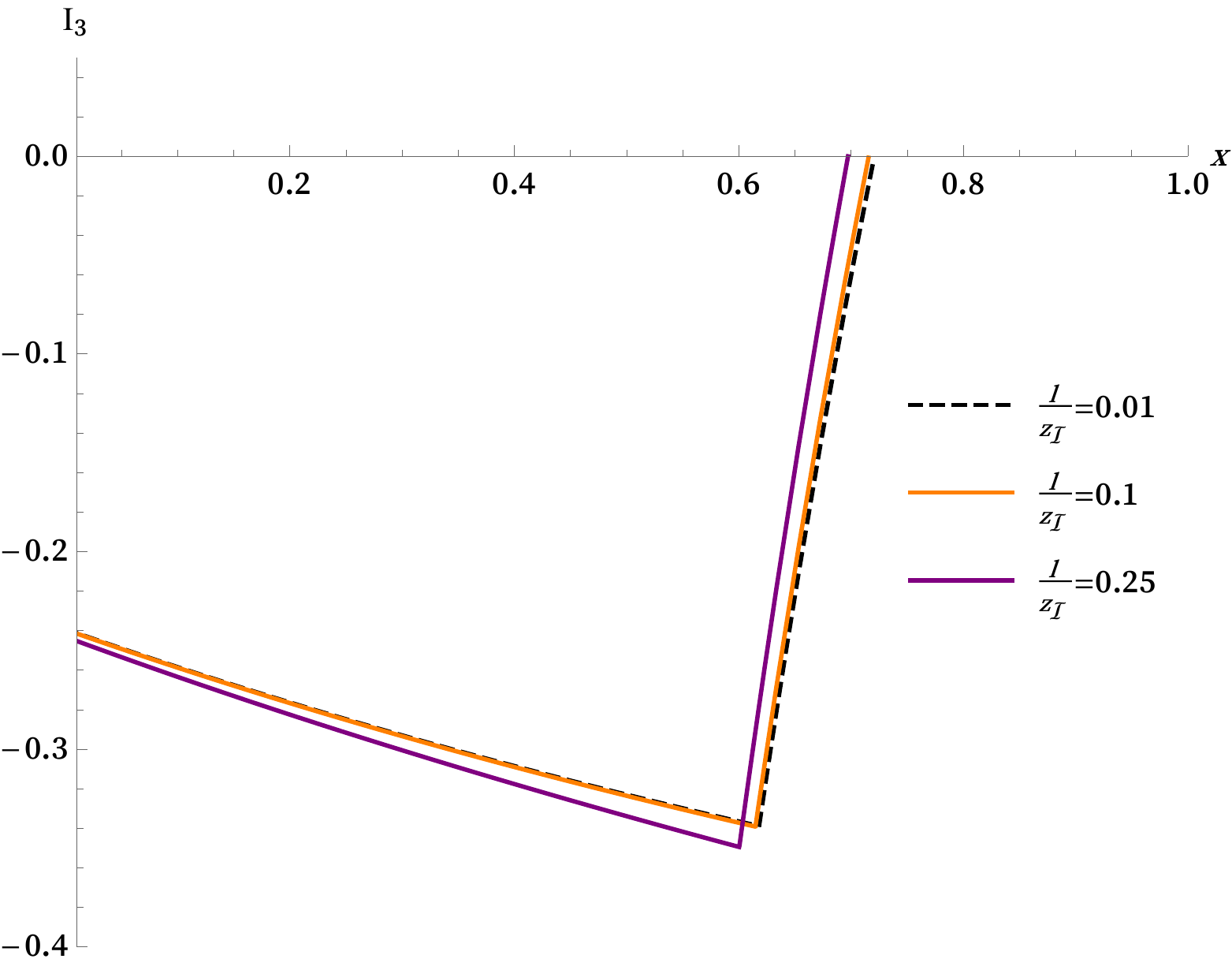}
		\caption{}
	\end{subfigure}
	\caption{Holographic tripartite information for three strips of equal widths and equal separation lengths: (a) effect of including higher orders in perturbation series, (b) tripartite information for different $\frac{\ell}{z_I}$ ratio. The $y$-axes in these plots are $\ell \times I_3(A: B: C)$ and we choose $\frac{L^2 \ell_2}{2 G_N^{(4)}} = 1$.}
	\label{tripartite}
\end{figure}

The holographic tripartite information is plotted against the dimensionless ratio $\frac{h}{\ell} \left(=x\right)$ in figure \ref{tripartite}. It is shown to satisfy the inequality \eqref{tripartite_inequality} and decreases with increase of charged excitation. We should note that the fulfilment of the inequality \eqref{tripartite_inequality} is a non-trivial check that the geometry \eqref{Lif_soln} could indeed be a classical holographic dual to a state of the strongly coupled Lifshitz QFT.

\section{Entanglement wedge cross section}\label{sec3}
In recent times, fair amount of attention has been devoted to the study of entanglement wedge cross-section (EWCS) in holographic theories \cite{Jeong:2019xdr, BabaeiVelni:2019pkw, BabaeiVelni:2020wfl, Boruch:2020wbe, Liu:2020blk, Jain:2020rbb, Amrahi:2021lgh, DiNunno:2021eyf, Sahraei:2021wqn, Chowdhury:2021idy, Ali-Akbari:2021zsm, ChowdhuryRoy:2022dgo, Vasli:2022kfu,Asadi:2022mvo}. The geometric quantity is usually defined as
\begin{equation} \label{EWCS_defn}
	E_{W} \left(A, B\right) = \frac{\text{Area}\left(\Sigma_{AB}\right)}{4 G_N},
\end{equation}
where $\Sigma_{AB}$ refers to the minimum cross-section of the entanglement wedge \cite{Wall:2012uf, Czech:2012bh, Headrick:2014cta} of the boundary subsystem $A \cup B$. Among other things, the EWCS has received attention because of it being thought of as a potential measure for correlation between $A$ and $B$ even for a mixed state \cite{Takayanagi:2017knl, Nguyen:2017yqw, Umemoto:2019jlz}. There are several proposals for the correct dual quantity to the EWCS on the field theory side, and they include: entanglement of purification \cite{Takayanagi:2017knl, Nguyen:2017yqw}, entanglement negativity \cite{Kudler-Flam:2018qjo, Kusuki:2019zsp}, odd entropy \cite{Tamaoka:2018ned} and reflected entropy \cite{Dutta:2019gen}.
\begin{figure}[t]
	\centering
	\includegraphics[scale = 0.2]{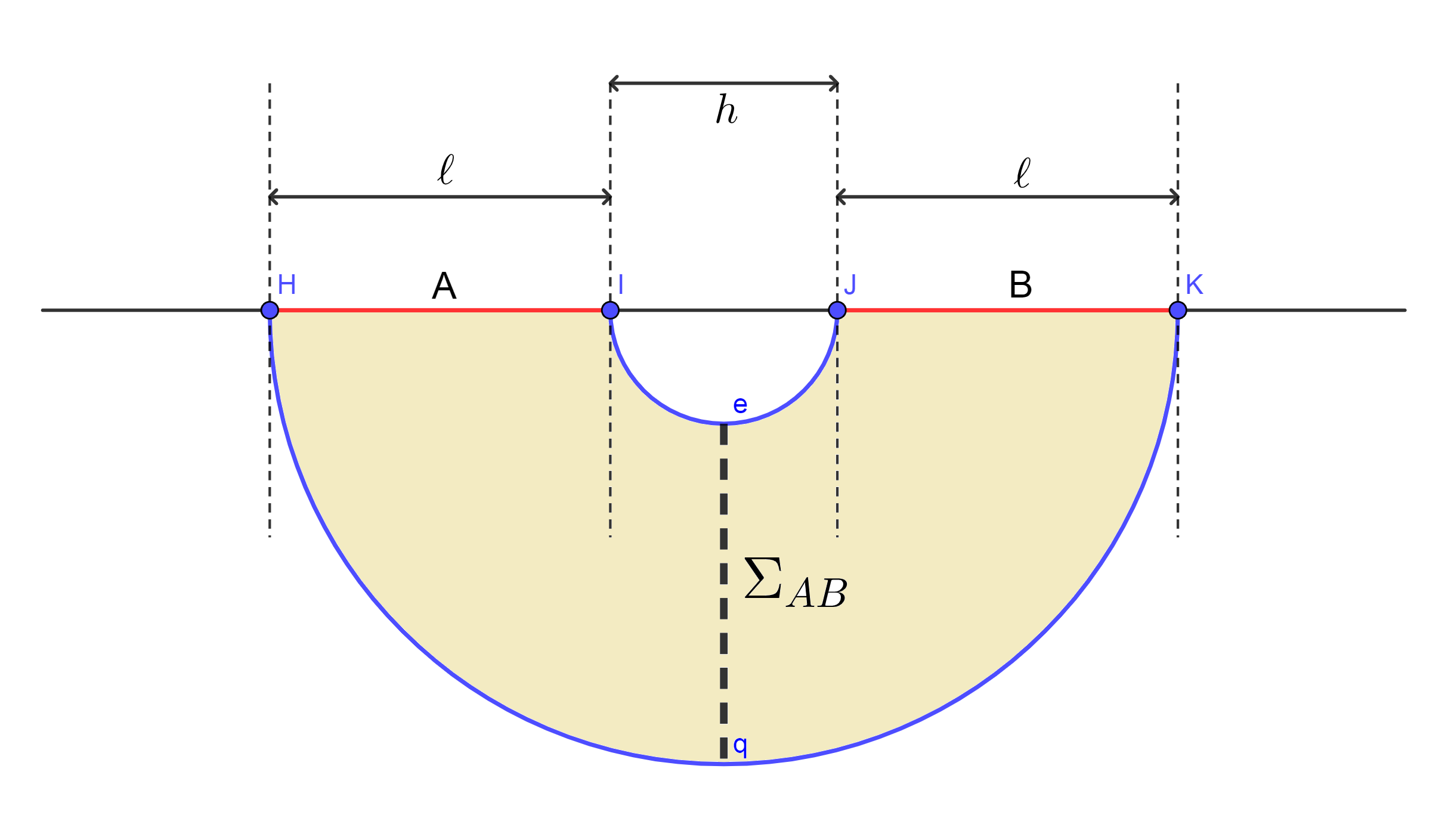}
	\caption{The entanglement wedge cross-section $\left(\Sigma_{AB}\right)$ for a symmetric configuration of two strips of width $\ell$ and separation length $h$. We only illustrate the connected phase as $\Sigma_{AB}$ vanishes otherwise.}
	\label{fig_entanglement_wedge}
\end{figure}

It is the easiest to calculate the EWCS for a configuration when the boundary subsystem constitutes of two disjoint strips of equal width $\ell$. As has been mentioned repeatedly, the entanglement wedge for such a configuration can either be connected or disconnected. In the disconnected phase, EWCS is identically zero because there doesn't exist a cross-section to begin with. In the connected phase, symmetry considerations help us decide that $\Sigma_{AB}$ would run along the radial direction and is the co-dimension 2 hypersurface in the bulk that connects the two turning points $z_{\ast}\left(\ell\right)$ and $z_{\ast}\left(h + 2\ell\right)$, with $h$ being the separation between the disjoint parts, see figure \ref{fig_entanglement_wedge}. Its area may be determined by minimizing the integral
\begin{equation} \label{EWCS_intgrl}
\begin{split}
\mathcal{A}_{\Sigma} &= 2L^2\ell_2\int_{z_{\ast}(h)}^{z_{\ast}(h + 2\ell)} \frac{dz}{z^2}\, h(z)^{\frac{1}{2}},\\
&= 2L^2\ell_2 \left(\int_{\epsilon}^{z_{\ast}(h + 2\ell)} \frac{dz}{z^2}\, h(z)^{\frac{1}{2}} - \int_{\epsilon}^{z_{\ast}(h)} \frac{dz}{z^2}\, h(z)^{\frac{1}{2}} \right),
\end{split}	
\end{equation} 
the integrand above can be easily read off from \eqref{HEE_intgrl} by putting $x_1'(z) = 0$. Let us remind ourselves that the integral diverges near the boundary where $z \to 0$, an ultraviolet cutoff $\epsilon$ is put to regulate the divergence.

We adopt the same perturbation series technique used to obtain the HEE in \cite{Maulik:2019qup}. We assume that the strip-width is very small so that the Ryu-Takayanagi hypersurfaces do not penetrate deep into the bulk. Then, we may write a series expansion of the above integrand in powers of $\frac{z_{\ast}}{z_I}$, i.e.
\begin{equation}
	\int_{\epsilon}^{z_{\ast}} \frac{dz}{z^2}\, h(z)^{\frac{1}{2}} \approx \int_{\epsilon}^{z_{\ast}} \frac{dz}{z^2} \left(1 + \frac{1}{2}\left(\frac{z_{\ast}}{z_I}\right)^2 z^2 - \frac{1}{8}\left(\frac{z_{\ast}}{z_I}\right)^4 z^4 + \mathcal{O}\left(\frac{z_{\ast}^6}{z_I^6}\right)\right),
\end{equation}
the first term in the parentheses would give a divergent result after integration, but the rest of the term remains finite as $\epsilon \to 0$. The above can be very easily integrated to obtain
\begin{equation}
\int_{\epsilon}^{z_{\ast}} \frac{dz}{z^2}\, h(z)^{\frac{1}{2}} = \frac{1}{\epsilon} + \frac{1}{z_{\ast}}\left( -1 + \frac{1}{2}\frac{z_{\ast}^2}{z_{I}^2} - \frac{1}{24}\frac{z_{\ast}^4}{z_{I}^4}\right).
\end{equation} 
We can then use eq. \eqref{turnpt} to write the turning point as $z_{\ast} = z_{\ast}\left(\ell\right)$. By substitution in the original formula \eqref{EWCS_intgrl} we finally obtain
\begin{multline}\label{EWCS_final}
	4 G_N^{(4)} \times E_{W}\left(A, B\right) = \frac{2 L^2 \ell_2}{\ell} \left(-2 b_0 \left(\frac{1}{x + 2} - \frac{1}{x} \right) + \frac{1}{4} \left(\frac{\ell}{z_I}\right)^2 \left(\left(x + 2\right) - x\right) \left(\frac{1}{b_0} - \frac{I_1}{b_0^2}\right) \right.\\ \left. - 
	\frac{1}{192} \left(\frac{\ell}{z_I}\right)^4 \left(\left(x + 2\right)^3 - x^3 \right) \left(1 + 6\frac{I_1}{b_0} - 12 \frac{I_1^2}{b_0^2} - 3 \frac{I_2}{b_0} \right) \frac{1}{b_0^3}\right),
\end{multline}
where as before, $x$ denotes the ratio $\frac{h}{\ell}$ and the constants $b_0, I_1$, and $I_2$ are the same as in eq. \eqref{beta_fn_parameters}. The divergent contribution from two parts once again cancel each other and render a finite quantity.
\begin{figure}[t]
	\begin{subfigure}{0.48\textwidth}
		\centering
		\includegraphics[width=\textwidth]{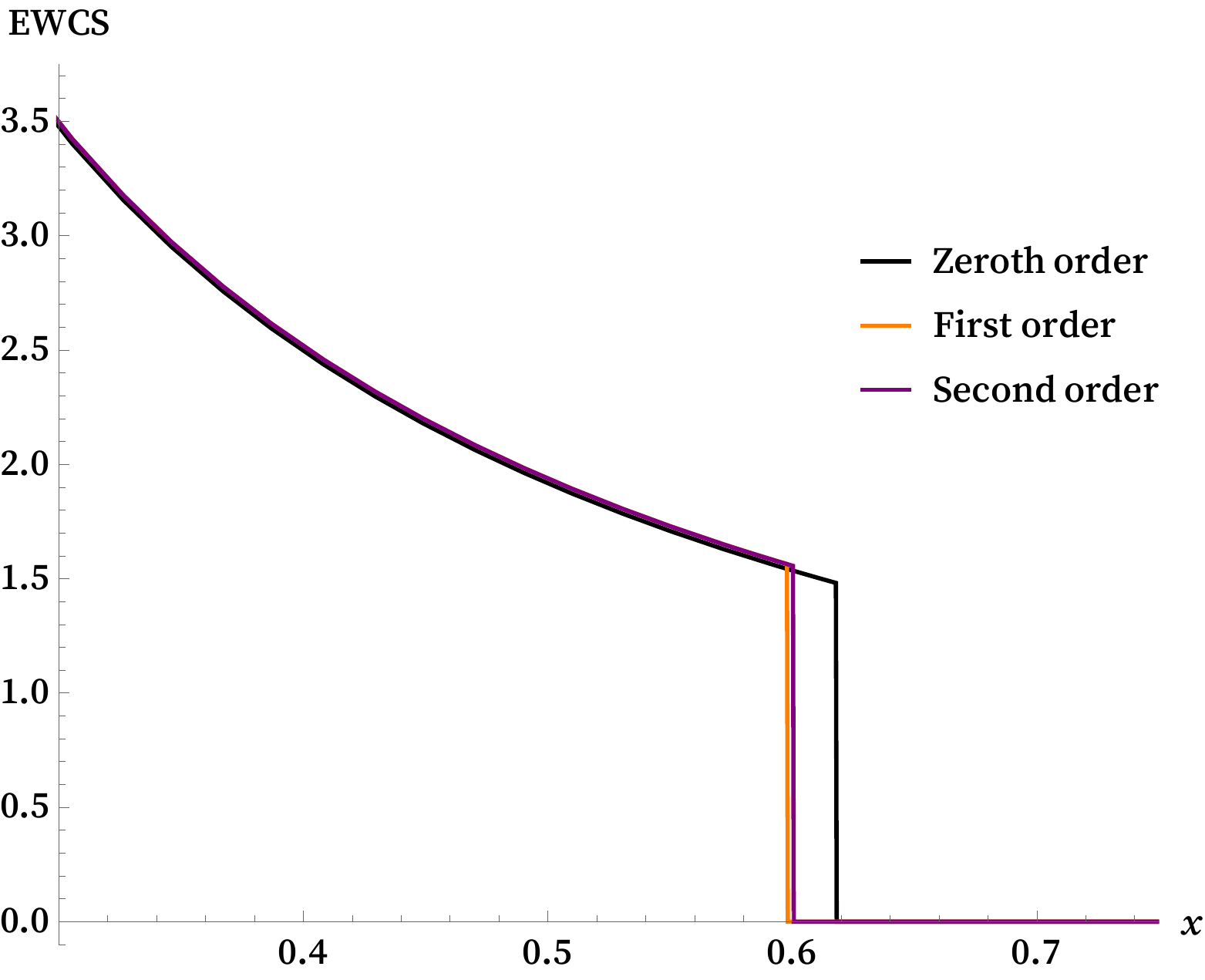}
		\caption{}
	\end{subfigure}
	\hfill
	\begin{subfigure}{0.48\textwidth}
		\centering
		\includegraphics[width=\textwidth]{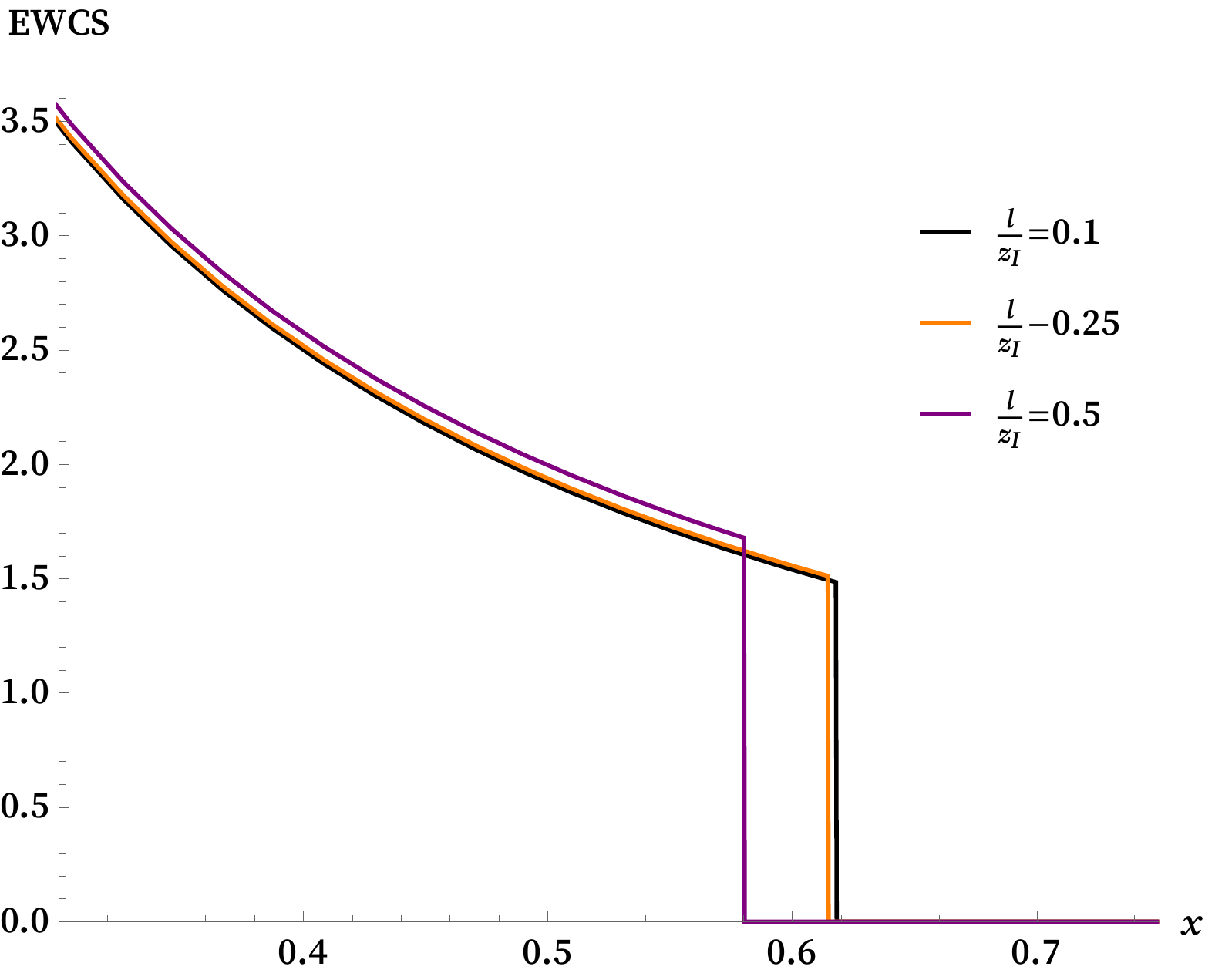}
		\caption{}
	\end{subfigure}
	\caption{Entanglement wedge cross-section (a) at different orders for $\frac{\ell}{z_I} = 0.25$, and (b) for different $\frac{\ell}{z_I}$ ratio. Plots are in units where $\frac{L^2 \ell_2}{2 G_N^{(4)}} = 1$.}
	\label{EWCS_fig}
\end{figure}

We plot the change of EWCS (or rather $\ell$ times the EWCS) in figure \ref{EWCS_fig}. Similar to mutual information, EWCS starts from a positive value when the entanglement wedge is connected and monotonically decreases as the separation between the two strips is increased. At the critical separation length (which is the solution to eq. \eqref{critical_length}) the entanglement wedge gets disconnected. Therefore, the EWCS goes to zero and remain so, thus it displays a discontinuity at the transition point.

We also observe that as the ratio $\frac{\ell}{z_I}$ increases, the value of EWCS for the same $x$ also increases. This is opposite to the previous observation about mutual information. Thus if $E_W$ is indeed a measure of mixed state entanglement, then we are led to the conclusion that the entanglement between $A$ and $B$ actually increases with increasing amount of excitation.

A comparison of the figures \ref{MI_eq} and \ref{EWCS_fig} also reveals that the inequality \cite{Nguyen:2017yqw}
\begin{equation}
	E_W\left(A, B\right) \geq \frac{I\left(A:B\right)}{2},
\end{equation}
is always obeyed for our non-relativistic system.

\section{Comparison with relativistic CFT} \label{sec_comparison}
So far we have been discussing the properties of entanglement in an excited state of a non-relativistic Lifshitz quantum field theory. In this section we compare our results for mutual and tripartite information, and entanglement wedge cross-section with similar quantities obtained in a relativistic CFT. We shall see that at the outset, the entanglement properties of the two theories are very similar. MI and EWCS in relativistic CFTs was studied extensively in \cite{Fischler:2012uv, BabaeiVelni:2019pkw, Sahraei:2021wqn}. For brevity, we re-derive some of their results.

We consider a $\left(3+1\right)$-dimensional AdS black hole as the gravity dual to a $\left(2+1\right)$-dimensional CFT at a finite temperature. The space-time metric is
\begin{equation} \label{AdS4_metric}
	ds^2 = \frac{L^2}{z^2}\left(-f(z)dt^2 + \frac{dz^2}{f(z)} + dx_1^2 + dx_2^2 \right), \quad f(z) = 1 - \frac{z^3}{z_H^3},
\end{equation}
where $z_H$ denotes the position of the event horizon. The temperature of the boundary CFT is equal to the black hole temperature, and is given by $T_H = \frac{3}{4\pi z_H}$.

We are interested in perturbative change in entanglement entropy due to finite temperature excitation. To achieve this, we assume the geometry \eqref{AdS4_metric} as a perturbation over pure AdS space-time and expand the blackening function $f(z)$ in a Taylor series, keeping terms up to $\mathcal{O}\left(z_H^{-6}\right)$. Obviously, the perturbative approximation is only valid as long as $\ell 
\ll z_H$. Then it is straightforward to apply the Ryu-Takayanagi proposal and obtain the entanglement entropy for a strip: $\{x_1 \in \left(-\frac{\ell}{2}, \frac{\ell}{2}\right),\; x_2 \in \left(0, \ell_2\right) \}$ in the above geometry \cite{Mishra:2015cpa, Maulik:2020tzm}
\begin{equation}
	S_E^{\text{rel}} = \frac{L^2 \ell_2}{2 G_N^{(4)}}\left(\frac{1}{\epsilon} - \frac{2 b_0'^2}{\ell} + \frac{\ell^2}{z_H^3} \frac{b_1'}{4 b_0'^2} - \frac{\ell^5}{z_H^6}\left(\frac{1}{64}\frac{b_1'^2}{b_0'^6} - \frac{3}{320}\frac{b_2'}{b_0'^5} \right) \right),
\end{equation}
where the coefficients are results of integration expressed in terms of beta functions
\begin{equation}
	b_0' = \frac{B\left(\frac{3}{4}, \frac{1}{2}\right)}{4},
	\quad b_1' = \frac{B\left(\frac{3}{2}, \frac{1}{2}\right)}{4},
	\quad b_2' = \frac{B\left(\frac{9}{4}, \frac{1}{2}\right)}{4},
\end{equation}
and $\epsilon$ is an ultraviolet cutoff as usual. The turning point $z_{\ast}$ in the relativistic case is related to the strip-width $\ell$ by
\begin{equation}
	z_{\ast} = \frac{\frac{\ell}{2 b_0'}}{1+\frac{1}{2}\frac{\ell^3}{z_H^3}\frac{b_1'}{8 b_0'^4}+\frac{\ell^6}{z_H^6}\left(\frac{3}{512}\frac{b_2'}{b_0'^7}-\frac{3}{256}\frac{b_1'^2}{b_0'^8} \right)},
\end{equation}

The mutual information between two strips of equal widths in the relativistic set-up would be given by (as before we denote $\frac{h}{\ell}$ by $x$)
\begin{align}
	I^{\text{rel}}\left(A:B\right) = \begin{dcases*}
		\mathcal{I}^{\text{rel}}_{AB}, & if $ x < x_c $, \\
		0, & if $ x > x_c $,
	\end{dcases*}
\end{align}
where $\mathcal{I}^{\text{rel}}_{AB}$ is
\begin{equation}
	\begin{split}
	\mathcal{I}^{\text{rel}}_{AB} = \frac{L^2 \ell_2}{2 G_N^{(4)}}\times \frac{1}{\ell} \left(\left(\frac{1}{x} + \frac{1}{x + 2} -2\right) 2b_0'^2 + \frac{\ell^3}{z_H^3}\left(2 - x^2 - \left(x+2\right)^2 \right)\frac{b_1'}{4 b_0'^2}\right.\\ \left. + \frac{\ell^6}{z_H^6} \left(x^5 + \left(x+2\right)^5 - 2 \right)\left(\frac{1}{64}\frac{b_1'^2}{b_0'^6} - \frac{3}{320}\frac{b_2'}{b_0'^5} \right) \right),
	\end{split}
\end{equation}
and the critical point $x_c$ can be obtained by solving for the root of
\begin{multline}
	\left(\frac{1}{x} + \frac{1}{x + 2} -2\right) + \frac{\ell^3}{z_H^3}\left(2 - x^2 - \left(x+2\right)^2 \right)\frac{b_1'}{8 b_0'^4}\\ + \frac{\ell^6}{z_H^6} \left(x^5 + \left(x+2\right)^5 - 2 \right)\left(\frac{1}{128}\frac{b_1'^2}{b_0'^8} - \frac{3}{640}\frac{b_2'}{b_0'^7} \right) = 0\,.
\end{multline}
One may compare the above equation with equation \eqref{critical_length} and see the difference between the two cases.

\begin{figure}[t]
	\begin{subfigure}{0.48\textwidth}
		\centering
		\includegraphics[width=\textwidth]{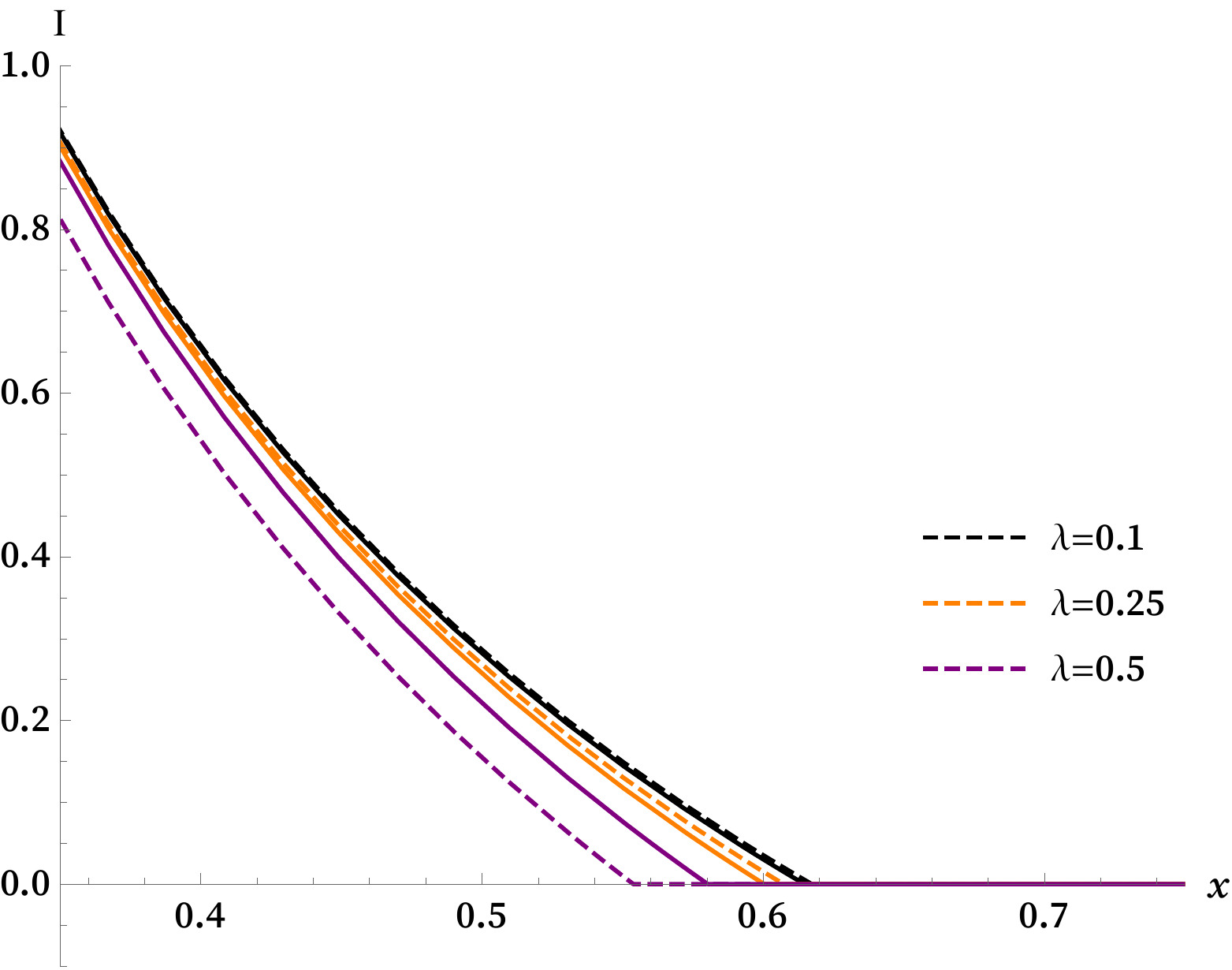}
		\caption{ }
		\label{fig_MI_comparison}
	\end{subfigure}
	\hfill
	\begin{subfigure}{0.48\textwidth}
		\centering
		\includegraphics[width=\textwidth]{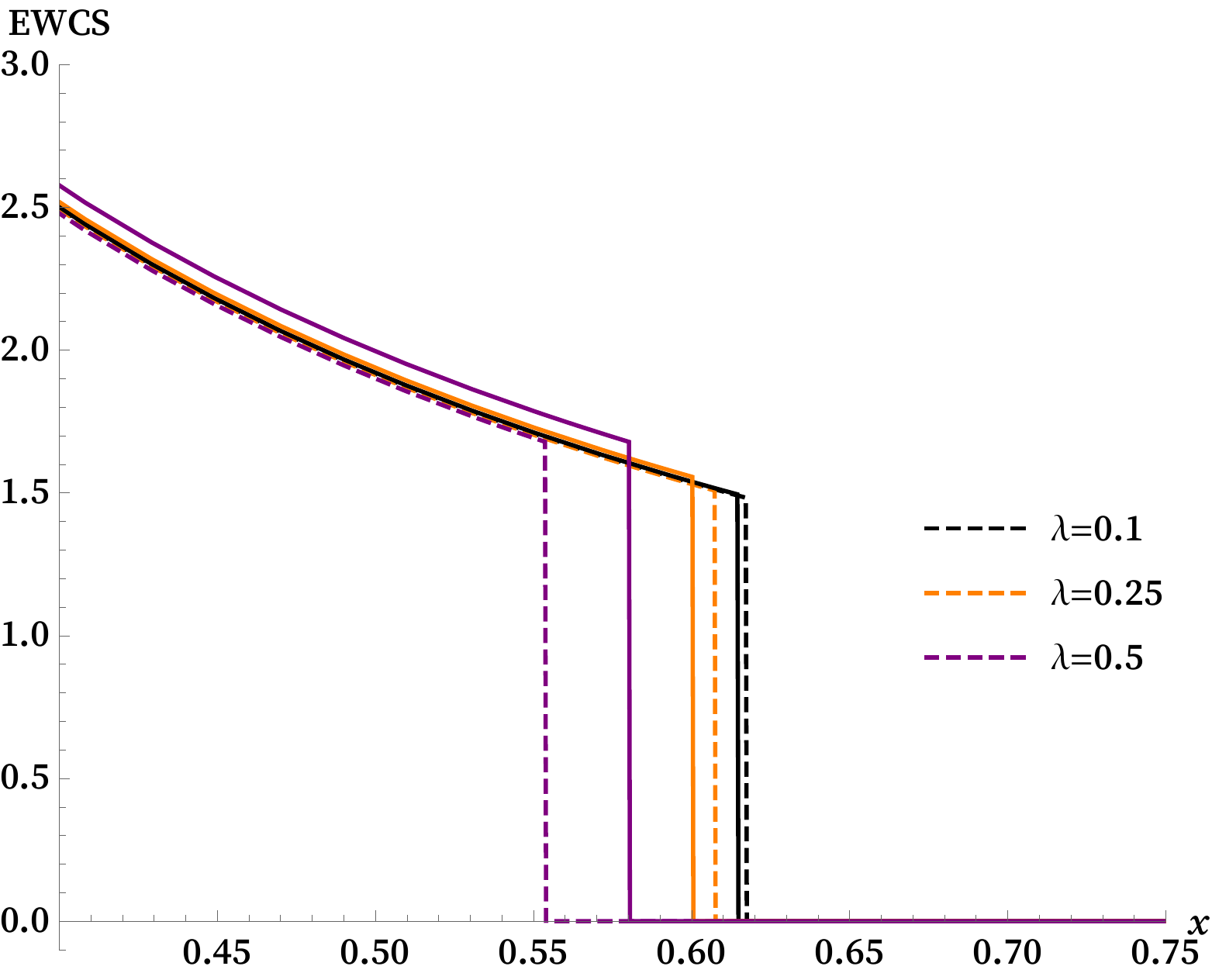}
		\caption{ }
		\label{fig_EWCS_comparison}
	\end{subfigure}
	\caption{Comparison between (a) holographic mutual information, and (b) entanglement wedge cross-section in relativistic and non-relativistic theories, the $Y$-axis is $\ell$ times the quantities. The degree of excitation is denoted by $\lambda$, which equals $\frac{\ell^2}{z_I^2}$ or $\frac{\ell^3}{z_H^3}$ depending on the theory. The relativistic CFT results are given by dashed curves. Plot is drawn in units where $\frac{L^2\ell_2}{2G_N^{(4)}}=1$.}
	\label{fig_comparison}
\end{figure}

The mutual information, or rather any entanglement measure is the same in the ground state of both the relativistic and Lifshitz field theories. This is easy to understand from the area functional in equation \eqref{HEE_intgrl}, which is identical to the one in pure AdS$_4$ when $h(z) = 1$. As excitation in the system is increased, the difference between the two kinds of theories begin to show up. We observe that the mutual information, as a function of $\frac{h}{\ell}$ in a relativistic system may be greater or less than that in a Lifshitz theory depending on the amount of excitation, see figure \ref{fig_MI_comparison}. The difference between the results also increases with increasing excitation.

In a similar fashion, one can obtain the entanglement wedge cross-section for the equal-width configuration in the connected phase \cite{BabaeiVelni:2019pkw, Sahraei:2021wqn}. We only quote the final expression
\begin{multline}
	E^{\text{rel}}_W(A, B) = \frac{2 L^2 \ell_2}{4 G_N^{(4)}}\times\frac{1}{\ell} \left(2b_0'\left(\frac{1}{x} - \frac{1}{x+2}\right) + \frac{\ell^3}{z_H^3}\left(\left(x+2\right)^2 - x^2\right)\left(\frac{1}{16\,b_0'^2} - \frac{b_1'}{8\,b_0'^3} \right)\right. \\ \left. + \frac{\ell^6}{z_H^6}\left(\left(x+2\right)^5 - x^5\right)\left(\frac{3}{1280\,b_0'^5} - \frac{b_1'}{128\,b_0'^6} - \frac{15\,b_2'}{1280\,b_0'^6} + \frac{3\,b_1'^2}{128\,b_0'^7} \right) \right)
\end{multline}

The behaviour of EWCS as a function of $\frac{h}{\ell}$ is compared between a relativistic CFT and a Lifshitz field theory in figure \ref{fig_EWCS_comparison}. In contrast to the case of mutual information, the entanglement wedge cross-section always seems to be higher for the non-relativistic theory within the domain of validity of the thin strip-width approximation; although the difference between points of transition for the two theories depends on the amount of excitation. Also, EWCS in the asymptotically AdS spacetime is always seen to decrease with increasing excitation while the same quantity in the Lifshitz spacetime is witnessed to display an opposite behaviour. Therefore, the entanglement wedge cross-section carries non-trivial information about the excited state and may be considered a good candidate to distinguish between the two kinds of systems.

\begin{figure}[t]
	\centering
	\includegraphics[width=0.52\linewidth]{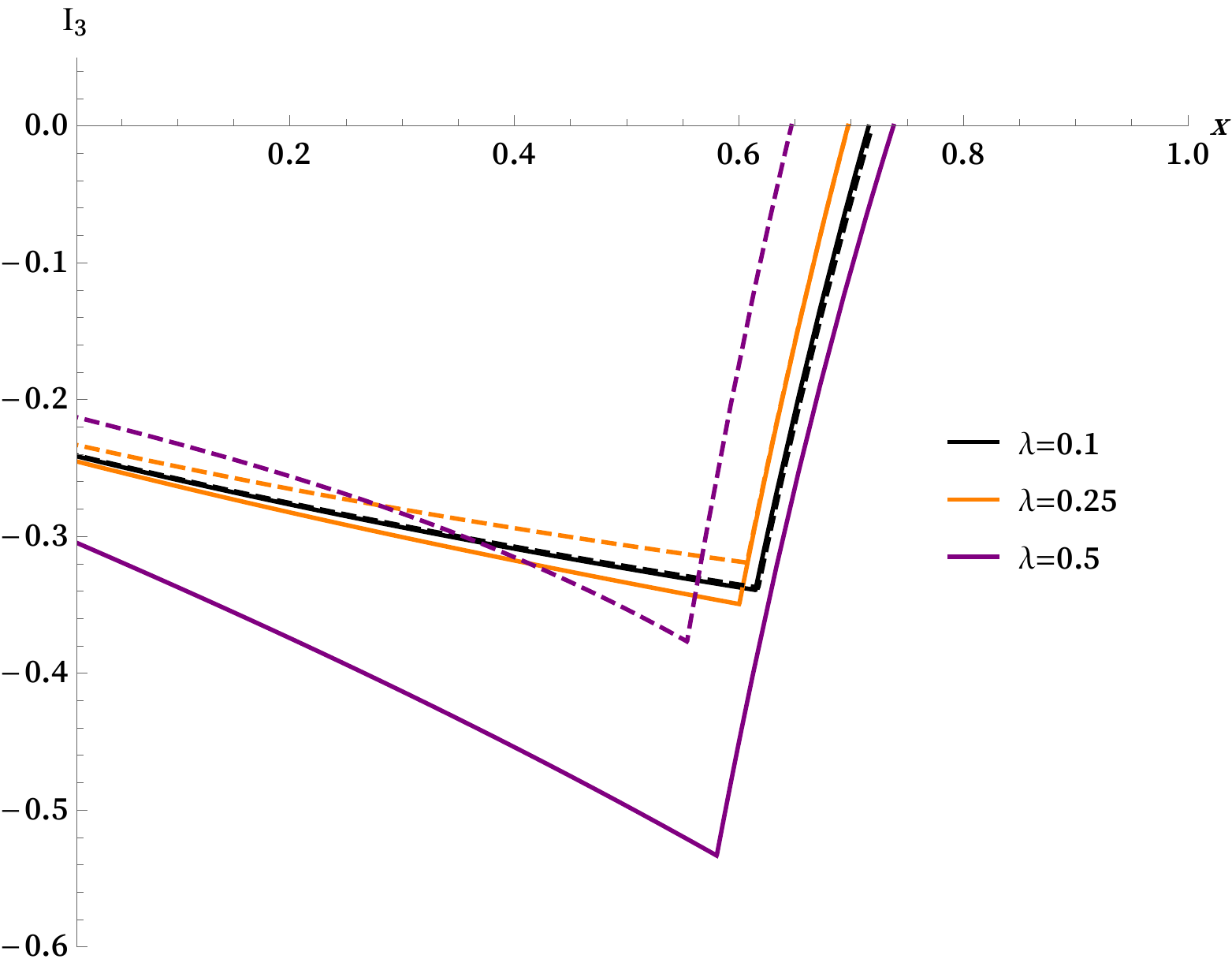}
	\caption{Comparison between holographic tripartite information $I_3\left(A:B:C\right)$in relativistic and non-relativistic theories, the $Y$-axis is $\ell$ times the quantity. The degree of excitation is denoted by $\lambda$, which equals $\frac{\ell^2}{z_I^2}$ or $\frac{\ell^3}{z_H^3}$ depending on the theory. The relativistic CFT results are given by dashed curves. Plots are drawn in units where $\frac{L^2\ell_2}{2G_N^{(4)}}=1$.}
	\label{fig_tpi_comparison}
\end{figure}

Finally, we compare the tripartite information $\left(I_{3}\right)$ for three disjoint strips of equal widths $\left(\ell\right)$ and separation $\left(h\right)$ between the two theories. The mathematical expressions in the relativistic theory are long and similar in principle to those one written in \ref{subsec_tpi} and we do not want to write them. Instead we directly refer to figure \ref{fig_tpi_comparison} for the main result. We observe that holographic tripartite information is usually more negative for the Lifshitz QFT, and the difference is more pronounced for increased excitations. For holographic theories, the tripartite information measures the `extensivity' and `monogamy' of mutual information \cite{Hayden:2011ag}. We conclude that mutual information between different parts of the subsystem in the dual excited state behaves less extensively in the non-relativistic theory. Moreover, in the non-relativistic theory the entanglement between different parts of the subsystem is less dominant over classical correlations in the non-relativistic Lifshitz quantum field theory.

\section{Concluding remarks}\label{sec4}
In this work we studied properties of quantum entanglement in an asymptotically Lifshitz spacetime, when the entangling subsystem in the dual QFT is made of more than one disjoint parts. Our studies were based on subsystems that had the geometry of a strip, and we assumed that the strip-width was very small compared to another relevant length-scale in the bulk to make analytic calculations tractable. To learn about the correlation and entanglement between the parts of the subregion, we studied three quantities: (i) mutual information, (ii) tripartite information, and (iii) entanglement wedge cross-section. The first two of them are defined using different linear combinations of entanglement entropy and happen to be free of any short-distance divergence. We looked at the geometric phase-transition of entanglement entropy for two boundary strips of equal and unequal widths within the range of validity of our perturbative calculation and pointed out the critical separation. By studying the mutual and tripartite information we found that the correlation between the parts of the subsystem always decreased when the degree of excitation in the system (measured by a smaller value of $z_I$) was increased. The tripartite information was also seen to obey the inequality \eqref{tripartite_inequality}, as expected for a holographic theory.

The entanglement wedge cross-section is a good measure of entanglement for mixed states. We studied the EWCS for two strips of equal width in the small-width approximation. It was observed to follow usual properties ascribed to it. Contrary to the holographic mutual information, however, the EWCS was seen to increase with increasing excitation in the system.

We also compared our results for the non-relativistic Lifshitz theory with similar quantities in a $\left(3+1\right)$-dimensional perturbed asymptotically AdS spacetime. The latter being dual to an excited state of a relativistic CFT. Although the source of excitation was completely different in both theories, since the relativistic CFT was thermal while the Lifshitz QFT was always at zero temperature; we were only interested in perturbative excitation infinitesimally close to the ground state and hence it made sense to compare between the two. The qualitative features in both theories were found to be similar. However, we noticed that the entanglement wedge cross-section remained larger for the Lifshitz case for a significant range of $\frac{h}{\ell}$ ratio; this implies that mixed state entanglement between parts of the non-relativistic theory is usually greater than in relativistic case. We also noticed that the change in EWCS with increase of excitations in the system is quite different for the relativistic and non-relativistic cases. We can, therefore, infer that the mixed state correlation between parts of the subsystem is behaving differently for the two theories. 

Despite the tremendous success of gauge/gravity duality, it still remains unclear whether it may be applied to any physical system amenable to low energy experiments. This work is a small step in the understanding of holography for non-relativistic field theories through the nature of entanglement. More works along related lines may open up a new door for studying strongly coupled quantum field theories. Having said that, it might be an interesting future direction to extend the analysis of this paper for other asymptotically Lifshitz spacetime, e.g. an $a=3$, $\theta=1$ theory as considered in \cite{Mishra:2018tzj}.

\section*{Acknowledgements}
The author acknowledges Harvendra Singh for useful discussions and a careful reading of the manuscript. The figures that depict the entanglement wedge were drawn using \emph{Geogebra} \cite{geogebra}.

\section*{Data availability statement}
Data sharing not applicable to this article as no datasets were generated or analysed during the current study.

\bibliographystyle{JHEP}
\bibliography{references}

\providecommand{\href}[2]{#2}\begingroup\raggedright\begin{thebibliography}{10}

\bibitem{Maldacena:1997re}
J.M.~Maldacena, \emph{{The Large N limit of superconformal field theories and
  supergravity}}, \href{https://doi.org/10.1023/A:1026654312961}{\emph{Int. J.
  Theor. Phys.} {\bfseries 38} (1999) 1113}
  [\href{https://arxiv.org/abs/hep-th/9711200}{{\ttfamily hep-th/9711200}}].

\bibitem{Witten:1998qj}
E.~Witten, \emph{{Anti-de Sitter space and holography}},
  \href{https://doi.org/10.4310/ATMP.1998.v2.n2.a2}{\emph{Adv. Theor. Math.
  Phys.} {\bfseries 2} (1998) 253}
  [\href{https://arxiv.org/abs/hep-th/9802150}{{\ttfamily hep-th/9802150}}].

\bibitem{Ryu:2006bv}
S.~Ryu and T.~Takayanagi, \emph{{Holographic derivation of entanglement entropy
  from AdS/CFT}},
  \href{https://doi.org/10.1103/PhysRevLett.96.181602}{\emph{Phys. Rev. Lett.}
  {\bfseries 96} (2006) 181602}
  [\href{https://arxiv.org/abs/hep-th/0603001}{{\ttfamily hep-th/0603001}}].

\bibitem{Ryu:2006ef}
S.~Ryu and T.~Takayanagi, \emph{{Aspects of Holographic Entanglement Entropy}},
  \href{https://doi.org/10.1088/1126-6708/2006/08/045}{\emph{JHEP} {\bfseries
  08} (2006) 045} [\href{https://arxiv.org/abs/hep-th/0605073}{{\ttfamily
  hep-th/0605073}}].

\bibitem{Hubeny:2007xt}
V.E.~Hubeny, M.~Rangamani and T.~Takayanagi, \emph{{A Covariant holographic
  entanglement entropy proposal}},
  \href{https://doi.org/10.1088/1126-6708/2007/07/062}{\emph{JHEP} {\bfseries
  07} (2007) 062} [\href{https://arxiv.org/abs/0705.0016}{{\ttfamily
  0705.0016}}].

\bibitem{VanRaamsdonk:2010pw}
M.~Van~Raamsdonk, \emph{{Building up spacetime with quantum entanglement}},
  \href{https://doi.org/10.1142/S0218271810018529}{\emph{Gen. Rel. Grav.}
  {\bfseries 42} (2010) 2323}
  [\href{https://arxiv.org/abs/1005.3035}{{\ttfamily 1005.3035}}].

\bibitem{Faulkner:2013ica}
T.~Faulkner, M.~Guica, T.~Hartman, R.C.~Myers and M.~Van~Raamsdonk,
  \emph{{Gravitation from Entanglement in Holographic CFTs}},
  \href{https://doi.org/10.1007/JHEP03(2014)051}{\emph{JHEP} {\bfseries 03}
  (2014) 051} [\href{https://arxiv.org/abs/1312.7856}{{\ttfamily 1312.7856}}].

\bibitem{Son:2008ye}
D.~Son, \emph{{Toward an AdS/cold atoms correspondence: A Geometric realization
  of the Schrodinger symmetry}},
  \href{https://doi.org/10.1103/PhysRevD.78.046003}{\emph{Phys. Rev. D}
  {\bfseries 78} (2008) 046003}
  [\href{https://arxiv.org/abs/0804.3972}{{\ttfamily 0804.3972}}].

\bibitem{Balasubramanian:2008dm}
K.~Balasubramanian and J.~McGreevy, \emph{{Gravity duals for non-relativistic
  CFTs}}, \href{https://doi.org/10.1103/PhysRevLett.101.061601}{\emph{Phys.
  Rev. Lett.} {\bfseries 101} (2008) 061601}
  [\href{https://arxiv.org/abs/0804.4053}{{\ttfamily 0804.4053}}].

\bibitem{Balasubramanian:2010uk}
K.~Balasubramanian and K.~Narayan, \emph{{Lifshitz spacetimes from AdS null and
  cosmological solutions}},
  \href{https://doi.org/10.1007/JHEP08(2010)014}{\emph{JHEP} {\bfseries 08}
  (2010) 014} [\href{https://arxiv.org/abs/1005.3291}{{\ttfamily 1005.3291}}].

\bibitem{Kachru:2008yh}
S.~Kachru, X.~Liu and M.~Mulligan, \emph{{Gravity duals of Lifshitz-like fixed
  points}}, \href{https://doi.org/10.1103/PhysRevD.78.106005}{\emph{Phys. Rev.
  D} {\bfseries 78} (2008) 106005}
  [\href{https://arxiv.org/abs/0808.1725}{{\ttfamily 0808.1725}}].

\bibitem{Maulik:2019qup}
S.~Maulik and H.~Singh, \emph{{Holographic entanglement entropy for
  $Lif_4^{(2)}\times {S}^1\times S^5$ spacetime with string excitations}},
  \href{https://doi.org/10.1103/PhysRevD.103.066003}{\emph{Phys. Rev. D}
  {\bfseries 103} (2021) 066003}
  [\href{https://arxiv.org/abs/1911.10865}{{\ttfamily 1911.10865}}].

\bibitem{Singh:2017wei}
H.~Singh, \emph{{D2-D8 system with massive strings and the Lifshitz
  spacetimes}}, \href{https://doi.org/10.1007/JHEP04(2017)011}{\emph{JHEP}
  {\bfseries 04} (2017) 011}
  [\href{https://arxiv.org/abs/1701.00968}{{\ttfamily 1701.00968}}].

\bibitem{Singh:2018ibp}
H.~Singh, \emph{{RG flows and cascades of $Lif_4^{(2)}\times S^1\times S^5$
  vacua}}, \href{https://doi.org/10.1007/JHEP09(2018)120}{\emph{JHEP}
  {\bfseries 09} (2018) 120}
  [\href{https://arxiv.org/abs/1805.05163}{{\ttfamily 1805.05163}}].

\bibitem{Takayanagi:2017knl}
T.~Takayanagi and K.~Umemoto, \emph{{Entanglement of purification through
  holographic duality}},
  \href{https://doi.org/10.1038/s41567-018-0075-2}{\emph{Nature Phys.}
  {\bfseries 14} (2018) 573}
  [\href{https://arxiv.org/abs/1708.09393}{{\ttfamily 1708.09393}}].

\bibitem{Nguyen:2017yqw}
P.~Nguyen, T.~Devakul, M.G.~Halbasch, M.P.~Zaletel and B.~Swingle,
  \emph{{Entanglement of purification: from spin chains to holography}},
  \href{https://doi.org/10.1007/JHEP01(2018)098}{\emph{JHEP} {\bfseries 01}
  (2018) 098} [\href{https://arxiv.org/abs/1709.07424}{{\ttfamily
  1709.07424}}].

\bibitem{Umemoto:2019jlz}
K.~Umemoto, \emph{{Quantum and Classical Correlations Inside the Entanglement
  Wedge}}, \href{https://doi.org/10.1103/PhysRevD.100.126021}{\emph{Phys. Rev.
  D} {\bfseries 100} (2019) 126021}
  [\href{https://arxiv.org/abs/1907.12555}{{\ttfamily 1907.12555}}].

\bibitem{Chakraborty:2014lfa}
S.~Chakraborty, P.~Dey, S.~Karar and S.~Roy, \emph{{Entanglement thermodynamics
  for an excited state of Lifshitz system}},
  \href{https://doi.org/10.1007/JHEP04(2015)133}{\emph{JHEP} {\bfseries 04}
  (2015) 133} [\href{https://arxiv.org/abs/1412.1276}{{\ttfamily 1412.1276}}].

\bibitem{MohammadiMozaffar:2017nri}
M.R.~Mohammadi~Mozaffar and A.~Mollabashi, \emph{{Entanglement in Lifshitz-type
  Quantum Field Theories}},
  \href{https://doi.org/10.1007/JHEP07(2017)120}{\emph{JHEP} {\bfseries 07}
  (2017) 120} [\href{https://arxiv.org/abs/1705.00483}{{\ttfamily
  1705.00483}}].

\bibitem{Parker:2017lnh}
D.E.~Parker, R.~Vasseur and J.E.~Moore, \emph{{Entanglement Entropy in Excited
  States of the Quantum Lifshitz Model}},
  \href{https://doi.org/10.1088/1751-8121/aa70b3}{\emph{J. Phys. A} {\bfseries
  50} (2017) 254003} [\href{https://arxiv.org/abs/1702.07433}{{\ttfamily
  1702.07433}}].

\bibitem{He:2017wla}
T.~He, J.M.~Magan and S.~Vandoren, \emph{{Entanglement Entropy in Lifshitz
  Theories}}, \href{https://doi.org/10.21468/SciPostPhys.3.5.034}{\emph{SciPost
  Phys.} {\bfseries 3} (2017) 034}
  [\href{https://arxiv.org/abs/1705.01147}{{\ttfamily 1705.01147}}].

\bibitem{Mishra:2018tzj}
R.~Mishra and H.~Singh, \emph{{Entanglement entropy at higher orders for the
  states of $a = 3$ Lifshitz theory}},
  \href{https://doi.org/10.1016/j.nuclphysb.2018.11.012}{\emph{Nucl. Phys. B}
  {\bfseries 938} (2019) 307}
  [\href{https://arxiv.org/abs/1804.01361}{{\ttfamily 1804.01361}}].

\bibitem{MohammadiMozaffar:2018vmk}
M.~Mohammadi~Mozaffar and A.~Mollabashi, \emph{{Entanglement Evolution in
  Lifshitz-type Scalar Theories}},
  \href{https://doi.org/10.1007/JHEP01(2019)137}{\emph{JHEP} {\bfseries 01}
  (2019) 137} [\href{https://arxiv.org/abs/1811.11470}{{\ttfamily
  1811.11470}}].

\bibitem{Angel-Ramelli:2019nji}
J.~Angel-Ramelli, V.G.M.~Puletti and L.~Thorlacius, \emph{{Entanglement Entropy
  in Generalised Quantum Lifshitz Models}},
  \href{https://doi.org/10.1007/JHEP08(2019)072}{\emph{JHEP} {\bfseries 08}
  (2019) 072} [\href{https://arxiv.org/abs/1906.08252}{{\ttfamily
  1906.08252}}].

\bibitem{Angel-Ramelli:2020xvd}
J.~Angel-Ramelli, \emph{{Entanglement Entropy of Excited States in the Quantum
  Lifshitz Model}}, \href{https://doi.org/10.1088/1742-5468/abcd35}{\emph{J.
  Stat. Mech.} {\bfseries 2101} (2021) 013102}
  [\href{https://arxiv.org/abs/2009.02283}{{\ttfamily 2009.02283}}].

\bibitem{Mozaffar:2021nex}
M.R.M.~Mozaffar and A.~Mollabashi, \emph{{Time scaling of entanglement in
  integrable scale-invariant theories}},
  \href{https://doi.org/10.1103/PhysRevResearch.4.L022010}{\emph{Phys. Rev.
  Res.} {\bfseries 4} (2022) L022010}
  [\href{https://arxiv.org/abs/2106.14700}{{\ttfamily 2106.14700}}].

\bibitem{Romans:1985tz}
L.J.~Romans, \emph{{Massive N=2a Supergravity in Ten-Dimensions}},
  \href{https://doi.org/10.1016/0370-2693(86)90375-8}{\emph{Phys. Lett. B}
  {\bfseries 169} (1986) 374}.

\bibitem{Headrick:2010zt}
M.~Headrick, \emph{{Entanglement Renyi entropies in holographic theories}},
  \href{https://doi.org/10.1103/PhysRevD.82.126010}{\emph{Phys. Rev. D}
  {\bfseries 82} (2010) 126010}
  [\href{https://arxiv.org/abs/1006.0047}{{\ttfamily 1006.0047}}].

\bibitem{Fischler:2012uv}
W.~Fischler, A.~Kundu and S.~Kundu, \emph{{Holographic Mutual Information at
  Finite Temperature}},
  \href{https://doi.org/10.1103/PhysRevD.87.126012}{\emph{Phys. Rev. D}
  {\bfseries 87} (2013) 126012}
  [\href{https://arxiv.org/abs/1212.4764}{{\ttfamily 1212.4764}}].

\bibitem{Ben-Ami:2014gsa}
O.~Ben-Ami, D.~Carmi and J.~Sonnenschein, \emph{{Holographic Entanglement
  Entropy of Multiple Strips}},
  \href{https://doi.org/10.1007/JHEP11(2014)144}{\emph{JHEP} {\bfseries 11}
  (2014) 144} [\href{https://arxiv.org/abs/1409.6305}{{\ttfamily 1409.6305}}].

\bibitem{Alishahiha:2014jxa}
M.~Alishahiha, M.R.~Mohammadi~Mozaffar and M.R.~Tanhayi, \emph{{On the Time
  Evolution of Holographic n-partite Information}},
  \href{https://doi.org/10.1007/JHEP09(2015)165}{\emph{JHEP} {\bfseries 09}
  (2015) 165} [\href{https://arxiv.org/abs/1406.7677}{{\ttfamily 1406.7677}}].

\bibitem{Tanhayi:2015cax}
M.R.~Tanhayi, \emph{{Thermalization of Mutual Information in Hyperscaling
  Violating Backgrounds}},
  \href{https://doi.org/10.1007/JHEP03(2016)202}{\emph{JHEP} {\bfseries 03}
  (2016) 202} [\href{https://arxiv.org/abs/1512.04104}{{\ttfamily
  1512.04104}}].

\bibitem{Mirabi:2016elb}
S.~Mirabi, M.R.~Tanhayi and R.~Vazirian, \emph{{On the Monogamy of Holographic
  $n$-partite Information}},
  \href{https://doi.org/10.1103/PhysRevD.93.104049}{\emph{Phys. Rev. D}
  {\bfseries 93} (2016) 104049}
  [\href{https://arxiv.org/abs/1603.00184}{{\ttfamily 1603.00184}}].

\bibitem{Casini:2008wt}
H.~Casini and M.~Huerta, \emph{{Remarks on the entanglement entropy for
  disconnected regions}},
  \href{https://doi.org/10.1088/1126-6708/2009/03/048}{\emph{JHEP} {\bfseries
  03} (2009) 048} [\href{https://arxiv.org/abs/0812.1773}{{\ttfamily
  0812.1773}}].

\bibitem{Hayden:2011ag}
P.~Hayden, M.~Headrick and A.~Maloney, \emph{{Holographic Mutual Information is
  Monogamous}}, \href{https://doi.org/10.1103/PhysRevD.87.046003}{\emph{Phys.
  Rev. D} {\bfseries 87} (2013) 046003}
  [\href{https://arxiv.org/abs/1107.2940}{{\ttfamily 1107.2940}}].

\bibitem{Jeong:2019xdr}
H.-S.~Jeong, K.-Y.~Kim and M.~Nishida, \emph{{Reflected Entropy and
  Entanglement Wedge Cross Section with the First Order Correction}},
  \href{https://doi.org/10.1007/JHEP12(2019)170}{\emph{JHEP} {\bfseries 12}
  (2019) 170} [\href{https://arxiv.org/abs/1909.02806}{{\ttfamily
  1909.02806}}].

\bibitem{BabaeiVelni:2019pkw}
K.~Babaei~Velni, M.R.~Mohammadi~Mozaffar and M.H.~Vahidinia, \emph{{Some
  Aspects of Entanglement Wedge Cross-Section}},
  \href{https://doi.org/10.1007/JHEP05(2019)200}{\emph{JHEP} {\bfseries 05}
  (2019) 200} [\href{https://arxiv.org/abs/1903.08490}{{\ttfamily
  1903.08490}}].

\bibitem{BabaeiVelni:2020wfl}
K.~Babaei~Velni, M.R.~Mohammadi~Mozaffar and M.H.~Vahidinia, \emph{{Evolution
  of entanglement wedge cross section following a global quench}},
  \href{https://doi.org/10.1007/JHEP08(2020)129}{\emph{JHEP} {\bfseries 08}
  (2020) 129} [\href{https://arxiv.org/abs/2005.05673}{{\ttfamily
  2005.05673}}].

\bibitem{Boruch:2020wbe}
J.~Boruch, \emph{{Entanglement wedge cross-section in shock wave geometries}},
  \href{https://doi.org/10.1007/JHEP07(2020)208}{\emph{JHEP} {\bfseries 07}
  (2020) 208} [\href{https://arxiv.org/abs/2006.10625}{{\ttfamily
  2006.10625}}].

\bibitem{Liu:2020blk}
P.~Liu and J.-P.~Wu, \emph{{Mixed state entanglement and thermal phase
  transitions}}, \href{https://doi.org/10.1103/PhysRevD.104.046017}{\emph{Phys.
  Rev. D} {\bfseries 104} (2021) 046017}
  [\href{https://arxiv.org/abs/2009.01529}{{\ttfamily 2009.01529}}].

\bibitem{Jain:2020rbb}
P.~Jain and S.~Mahapatra, \emph{{Mixed state entanglement measures as probe for
  confinement}}, \href{https://doi.org/10.1103/PhysRevD.102.126022}{\emph{Phys.
  Rev. D} {\bfseries 102} (2020) 126022}
  [\href{https://arxiv.org/abs/2010.07702}{{\ttfamily 2010.07702}}].

\bibitem{Amrahi:2021lgh}
B.~Amrahi, M.~Ali-Akbari and M.~Asadi, \emph{{Temperature dependence of
  entanglement of purification in the presence of a chemical potential}},
  \href{https://doi.org/10.1103/PhysRevD.103.086019}{\emph{Phys. Rev. D}
  {\bfseries 103} (2021) 086019}
  [\href{https://arxiv.org/abs/2101.03994}{{\ttfamily 2101.03994}}].

\bibitem{DiNunno:2021eyf}
B.S.~DiNunno, N.~Jokela, J.F.~Pedraza and A.~P\"onni, \emph{{Quantum
  information probes of charge fractionalization in large-$N$ gauge theories}},
  \href{https://doi.org/10.1007/JHEP05(2021)149}{\emph{JHEP} {\bfseries 05}
  (2021) 149} [\href{https://arxiv.org/abs/2101.11636}{{\ttfamily
  2101.11636}}].

\bibitem{Sahraei:2021wqn}
M.~Sahraei, M.J.~Vasli, M.R.M.~Mozaffar and K.B.~Velni, \emph{{Entanglement
  wedge cross section in holographic excited states}},
  \href{https://doi.org/10.1007/JHEP08(2021)038}{\emph{JHEP} {\bfseries 08}
  (2021) 038} [\href{https://arxiv.org/abs/2105.12476}{{\ttfamily
  2105.12476}}].

\bibitem{Chowdhury:2021idy}
A.R.~Chowdhury, A.~Saha and S.~Gangopadhyay, \emph{{Entanglement wedge
  cross-section for noncommutative Yang-Mills theory}},
  \href{https://doi.org/10.1007/JHEP02(2022)192}{\emph{JHEP} {\bfseries 02}
  (2022) 192} [\href{https://arxiv.org/abs/2106.04562}{{\ttfamily
  2106.04562}}].

\bibitem{Ali-Akbari:2021zsm}
M.~Ali-Akbari, M.~Asadi and B.~Amrahi, \emph{{Non-conformal behavior of
  holographic entanglement measures}},
  \href{https://doi.org/10.1007/JHEP04(2022)014}{\emph{JHEP} {\bfseries 04}
  (2022) 014} [\href{https://arxiv.org/abs/2112.02565}{{\ttfamily
  2112.02565}}].

\bibitem{ChowdhuryRoy:2022dgo}
A.~Chowdhury~Roy, A.~Saha and S.~Gangopadhyay, \emph{{Mixed state information
  theoretic measures in boosted black brane}},
  \href{https://arxiv.org/abs/2204.08012}{{\ttfamily 2204.08012}}.

\bibitem{Vasli:2022kfu}
M.J.~Vasli, M.R.~Mohammadi~Mozaffar, K.~Babaei~Velni and M.~Sahraei,
  \emph{{Holographic Study of Reflected Entropy in Anisotropic Theories}},
  \href{https://arxiv.org/abs/2207.14169}{{\ttfamily 2207.14169}}.

\bibitem{Asadi:2022mvo}
M.~Asadi, B.~Amrahi and H.~Eshaghi-Kenari, \emph{{Probing Phase Structure of
  Strongly Coupled Matter with Holographic Entanglement Measures}},
  \href{https://arxiv.org/abs/2209.01586}{{\ttfamily 2209.01586}}.

\bibitem{Wall:2012uf}
A.C.~Wall, \emph{{Maximin Surfaces, and the Strong Subadditivity of the
  Covariant Holographic Entanglement Entropy}},
  \href{https://doi.org/10.1088/0264-9381/31/22/225007}{\emph{Class. Quant.
  Grav.} {\bfseries 31} (2014) 225007}
  [\href{https://arxiv.org/abs/1211.3494}{{\ttfamily 1211.3494}}].

\bibitem{Czech:2012bh}
B.~Czech, J.L.~Karczmarek, F.~Nogueira and M.~Van~Raamsdonk, \emph{{The Gravity
  Dual of a Density Matrix}},
  \href{https://doi.org/10.1088/0264-9381/29/15/155009}{\emph{Class. Quant.
  Grav.} {\bfseries 29} (2012) 155009}
  [\href{https://arxiv.org/abs/1204.1330}{{\ttfamily 1204.1330}}].

\bibitem{Headrick:2014cta}
M.~Headrick, V.E.~Hubeny, A.~Lawrence and M.~Rangamani, \emph{{Causality \&
  holographic entanglement entropy}},
  \href{https://doi.org/10.1007/JHEP12(2014)162}{\emph{JHEP} {\bfseries 12}
  (2014) 162} [\href{https://arxiv.org/abs/1408.6300}{{\ttfamily 1408.6300}}].

\bibitem{Kudler-Flam:2018qjo}
J.~Kudler-Flam and S.~Ryu, \emph{{Entanglement negativity and minimal
  entanglement wedge cross sections in holographic theories}},
  \href{https://doi.org/10.1103/PhysRevD.99.106014}{\emph{Phys. Rev. D}
  {\bfseries 99} (2019) 106014}
  [\href{https://arxiv.org/abs/1808.00446}{{\ttfamily 1808.00446}}].

\bibitem{Kusuki:2019zsp}
Y.~Kusuki, J.~Kudler-Flam and S.~Ryu, \emph{{Derivation of holographic
  negativity in AdS$_3$/CFT$_2$}},
  \href{https://doi.org/10.1103/PhysRevLett.123.131603}{\emph{Phys. Rev. Lett.}
  {\bfseries 123} (2019) 131603}
  [\href{https://arxiv.org/abs/1907.07824}{{\ttfamily 1907.07824}}].

\bibitem{Tamaoka:2018ned}
K.~Tamaoka, \emph{{Entanglement Wedge Cross Section from the Dual Density
  Matrix}}, \href{https://doi.org/10.1103/PhysRevLett.122.141601}{\emph{Phys.
  Rev. Lett.} {\bfseries 122} (2019) 141601}
  [\href{https://arxiv.org/abs/1809.09109}{{\ttfamily 1809.09109}}].

\bibitem{Dutta:2019gen}
S.~Dutta and T.~Faulkner, \emph{{A canonical purification for the entanglement
  wedge cross-section}},
  \href{https://doi.org/10.1007/JHEP03(2021)178}{\emph{JHEP} {\bfseries 03}
  (2021) 178} [\href{https://arxiv.org/abs/1905.00577}{{\ttfamily
  1905.00577}}].

\bibitem{Mishra:2015cpa}
R.~Mishra and H.~Singh, \emph{{Perturbative entanglement thermodynamics for AdS
  spacetime: Renormalization}},
  \href{https://doi.org/10.1007/JHEP10(2015)129}{\emph{JHEP} {\bfseries 10}
  (2015) 129} [\href{https://arxiv.org/abs/1507.03836}{{\ttfamily
  1507.03836}}].

\bibitem{Maulik:2020tzm}
S.~Maulik and H.~Singh, \emph{{Entanglement entropy and the first law at third
  order for boosted black branes}},
  \href{https://doi.org/10.1007/JHEP04(2021)065}{\emph{JHEP} {\bfseries 04}
  (2021) 065} [\href{https://arxiv.org/abs/2012.09530}{{\ttfamily
  2012.09530}}].

\bibitem{geogebra}
``Geogebra.'' \url{https://www.geogebra.org/}.

\end{thebibliography}\endgroup



\providecommand{\href}[2]{#2}\begingroup\raggedright\endgroup

\end{document}